\documentclass[aps,prl,reprint]{revtex4-1}
\usepackage{blindtext}

\usepackage{amsmath,amssymb,amsthm,amscd,graphicx}
\usepackage{psfrag}

\usepackage[utf8]{inputenc}
\usepackage{graphicx}
\usepackage{color}
\usepackage{hyperref}
\hypersetup{colorlinks,allcolors=blue}
\usepackage{filecontents}

\usepackage{color}

\usepackage[T1]{fontenc}

\usepackage[utf8]{inputenc}
\usepackage{graphicx}
\usepackage{amsmath,amsthm,amssymb,physics,mathtools}
\usepackage{tikz,tikz-cd, dynkin-diagrams}
\usepackage{epigraph}
\usepackage{hhline}
\usepackage{caption}
\usepackage{subcaption}
\usepackage{ytableau}
\usepackage{natbib}

\theoremstyle{definition}

\newcommand{\bea}{\begin{eqnarray}}
\newcommand{\eea}{\end{eqnarray}}

\newcommand{\brc}[1]{\left(#1\right)}

\def\v{\v}

\def\12{\frac{1}{2}}
\newcommand{\be}{\begin{equation}}
\newcommand{\ee}{\end{equation}}
\newcommand{\ba}{\begin{aligned}}
	\newcommand{\ben}{\begin{eqnarray}\displaystyle}
	\newcommand{\een}{\end{eqnarray}}

\begin{document}

		\title{ One-loop corrections to near extremal Kerr thermodynamics\\ from semiclassical Virasoro blocks}
		\author{Paolo Arnaudo}
    \email{P.Arnaudo@soton.ac.uk}
            \affiliation{Mathematical Sciences and STAG Research Centre, University of Southampton, Highfield, Southampton SO17 1BJ, UK}
            \author{Giulio Bonelli}
            \email{bonelli@sissa.it}
            \author{Alessandro Tanzini}
            \email{tanzini@sissa.it}
            \affiliation{International School of Advanced Studies (SISSA), via Bonomea 265, 34136 Trieste, Italy,}\affiliation {INFN, Sezione di Trieste}
	    \affiliation{Institute for Geometry and         Physics, IGAP, via Beirut 2, 34136  Trieste, Italy}

		
\begin{abstract}
Abstract: We propose a method to perform an exact calculation of one-loop quantum corrections to black hole entropy in terms of Virasoro semiclassical blocks. We analyse in detail four-dimensional Kerr black hole and show that in the near-extremal limit a branch of long-lived modes arises. We prove that the contribution of these modes accounts for a $(s-1/2)\log T_{\text{Hawking}}$ correction to the entropy for massless particles of spin $s=1,2$. We show that in the full calculation performed in the exact Kerr background the leading contribution actually is sourced by the near-horizon region only, and as such has a universal validity for any asymptotic behavior at infinity.
\end{abstract}

		\maketitle
		

Quantum corrections to near-extremal black hole (BH) entropy have been shown to be necessary in order
    to describe their low-temperature regime 
    \cite{Preskill:1991tb}.
     Logarithmic corrections should indeed exist in any gravitational theory and 
     characterize its infra-red physics. These are expected to be determined solely by the low energy degrees of freedom
    and to be independent of the higher derivatives or massive spectrum corrections.
    On the other hand, they are not universal, as they depend on the specific low-energy spectrum of the theory.
    Therefore, once logarithmic corrections are fixed, any acceptable micro-state description of quantum black holes shall reproduce them in the infrared (IR).
   
    Actually, our best agnostic models of quantum black hole radiation are still semi-classical and, as it has been largely discussed in the literature 
    these log-corrections arise from the one-loop analysis of quantum gravity path-integral in the IR. 
    Precise tests of this conjecture have been performed for supersymmetric BPS black holes in the context of AdS/CFT correspondence, where a microscopic calculation of the BH entropy is available 
\cite{Banerjee:2010qc,Banerjee:2011jp,Sen:2012kpz,Sen:2012cj,Bhattacharyya:2012ye,PandoZayas:2019hdb,Benini:2019dyp,David:2023btq,GonzalezLezcano:2023uar,H:2023qko}.

    Recent progresses in the challenging non-supersymmetric case have been obtained by analyzing the near-horizon BH geometry in terms of Schwarzian modes that describe its reparameterization symmetry \cite{Iliesiu:2020qvm,Kapec:2023ruw,Rakic:2023vhv,Banerjee:2023gll,Maulik:2024dwq}. In order to confirm these results it is clearly important to go beyond the near-horizon approximation and perform  
    an analysis of the full BH background. This has been accomplished recently for the BTZ
    case in three dimensions in \cite{Kapec:2024zdj} and in \cite{Kolanowski:2024zrq}, where
    numerical evidence has been found also for the Reissner-Nordstr\"om AdS$_4$ case.

    In \cite{PhysRevD.110.106006} we proposed an exact formula for the one-loop effective action on BH backgrounds with enough symmetries such that the BH perturbations admit a separation of variables leading to ordinary differential operators. In these cases, the relevant determinants can be computed by a generalisation of the Gelfand-Yaglom (GY) formula \cite{Dunne:2007rt}.
    We can show that, when the involved differential operators display only regular singularities, these formulae reduce to the ones conjectured by Denef-Hartnoll-Sachdev (DHS) in \cite{Denef:2009kn} based on analytic properties of the effective action of Euclidean quantum gravity. In particular, this conjecture implies that the one-loop determinants have zeroes at the BH (anti)-quasinormal modes (QNM). 
     The DHS formula has been the main tool to obtain the results in \cite{Kapec:2024zdj} where it has been proposed that      
     the logarithmic corrections 
     are due to
      particular modes in the black hole QNM spectrum developing a parametrically small negative imaginary part, called
      zero damping modes (ZDMs).

In this letter, we extend the formalism of \cite{PhysRevD.110.106006} to Kerr BH and we provide an exact formula
for the corresponding one-loop effective action. We notice that in this case the analytic structure of the determinant is more complicated than the DHS one, due to the presence of a branch cut in the frequency plane, see \eqref{detQNMs}. The exact calculation of QNM for Kerr BH 
\cite{Aminov:2020yma, Bonelli:2021uvf} can be performed in terms of semiclassical (confluent) Virasoro conformal blocks, whose explicit expression
can be computed from supersymmetric gauge theories  \cite{Alday:2009aq}. 
In particular, we show how to isolate the contribution of the ZDMs sector in the near-extremal limit and analyze the 
regularity conditions of the associated Euclidean wave functions.

By evaluating the contribution of these modes to the GY formula for the one-loop determinant, we find that the leading temperature factor 
for the massless spin $s=1,2$ mode
is
$T_H^{s-1/2}$, as conjectured in \cite{Iliesiu:2020qvm,Banerjee:2023quv}.
Our results are obtained by making use of explicit connection formulae for the Heun function given by successive analytic continuations ending in the near-horizon region. As we find that the origin of the ZDMs
is only due to the contribution in the near-horizon patch, this proves the universality of the mechanism, which
turns out to be independent of the geometry far away from the horizon. 

It turns out that the ZDMs contributing to the 
leading log-correction in the temperature are the ones with azimuthal quantum number $m=0$.
These modes, appearing in \eqref{MatsubaraKerr} for $m=0$, should correspond to the extension to the full Kerr geometry of the ones analysed in \cite{Kapec:2023ruw} in the near-horizon extremal Kerr throat approximation. It would be interesting to verify this in detail.

This separation of scales arises naturally from the Heun connection problem and is a characteristic feature of BH perturbation theory manifesting itself also for other physical quantities as scattering amplitudes and tidal deformations \cite{Bautista:2023sdf}.

The QNM of other BH
geometries have been studied with similar techniques in \cite{Novaes:2018fry,CarneirodaCunha:2019tia,Bianchi:2021xpr,Fioravanti:2021dce,Aminov:2023jve,Aminov:2024mul}. In particular, the analysis of ZDMs for Kerr-Newman in $\mathrm{(A)dS}_4$ and analogous higher-dimensional BH can be analyzed with similar techniques applied here. More in general, our results point out that ZDMs generically arise when considering confluences of second-order Fuchsian differential operators. 
Let us notice that log-corrections are reliable sub-leading terms 
in the black hole entropy only before the temperature becomes too low and non-perturbative corrections to the path integral become important  
\cite{Iliesiu:2022onk}.
At higher temperatures, instead, UV generated contributions picture shall dominate over the classical result. 
The relevant mechanisms at scales at which these effects happen are model-dependent and their description is well beyond the scope of this letter.
Our analysis is limited to the cases $s=1$ and $s=2$.
To extend it to massless $s=0$, one should refine the analysis of the ZDM contribution to next order in the temperature.
Moreover, it would also be interesting to analyze with similar techniques the contribution of fermions.
This should not be obtained by naively setting $s$ to half-integer values in our formulae but requires further analysis. It would moreover allow also to investigate the expected absence of logarithmic corrections in the supersymmetric case by studying the relevant BH geometries.


\vspace{0.5cm}

The four-dimensional asymptotically flat Kerr black hole metric in the 
Boyer-Lindquist coordinates is
\begin{equation}
\begin{aligned}
\mathrm{d}s^2=\,&-\mathrm{d}t^2+\mathrm{d}r^2+2\,a_{\text{BH}}\,\sin^2\theta\,\mathrm{d}r\,\mathrm{d}\phi\\
&+\left(r^2+a_{\text{BH}}^2\,\cos^2\theta\right)\,\mathrm{d}\theta^2+\left(r^2+a_{\text{BH}}^2\right)\,\sin^2\theta\,\mathrm{d}\phi^2\\
&+\frac{2\,M\,r}{r^2+a_{\text{BH}}^2\,\cos^2\theta}\left(\mathrm{d}t+\mathrm{d}r+a_{\text{BH}}\,\sin^2\theta\,\mathrm{d}\phi\right)^2,
\end{aligned}
\end{equation}
where $M$ is the mass of the black hole and $a_{\text{BH}}$ is the parameter describing its angular momentum.
The radial geometry admits two horizons: the event horizon $R_h=M+\sqrt{M^2-a_{\text{BH}}^2}$ and an inner Cauchy horizon $R_i=M-\sqrt{M^2-a_{\text{BH}}^2}$. 
The temperature and the angular velocity at the event horizon read
\begin{equation}
\begin{aligned}
T_H=\frac{R_h-R_i}{8 \pi  M R_h},\quad\quad
\Omega_H=\frac{a_{\text{BH}}}{2 M R_h}.
\end{aligned}
\end{equation}
Based on the Newman-Penrose formalism, Teukolsky \cite{PhysRevLett.29.1114} showed that in terms of curvature invariants, the perturbation equations decouple and separate for all Petrov type-D spacetimes.
Let us consider the Fourier-transform of a spin-$s$ field $\Phi(t,r,\theta.\phi)$ and expand it in spin-weighted spheroidal harmonics as
\begin{equation}
\Phi=\frac{1}{2\pi}\int e^{-i\,\omega\,t}\sum_{\ell=|s|}^{\infty}\sum_{m=-\ell}^{\ell}e^{i\,m\,\phi}{}_sS_{\ell\,m}(\theta,\phi)R_{\ell\,m}(r)\,\mathrm{d}\omega,
\end{equation}
where $\omega$ is the frequency of the perturbation, and $\ell$ and $m$ are the angular quantum numbers.
With the above separation of variables, both the radial and angular equations can be rewritten as confluent Heun equations \cite{Bonelli:2021uvf}:
\begin{equation}\label{confluentHeun}
\begin{aligned}
\mathcal{D}\,\psi\equiv
\psi''(z)+\biggl[&\frac{u-\frac{1}{2}+a_0^2+a_1^2}{z(z-1)}+\frac{\frac{1}{4}-a_1^2}{(z-1)^2}\\
&+\frac{\frac{1}{4}-a_0^2}{z^2}+\frac{\mu\,\epsilon}{z}-\frac{\epsilon^2}{4}\biggr]\psi(z)=0.
\end{aligned}
\end{equation}
In particular, we are interested in the dictionary for the radial equation, given by
\begin{equation}\label{dictionarykerr}
\begin{aligned}
z&=\frac{r-R_i}{R_h-R_i},\ \ \psi(z)=(r^2-2Mr+a_{\text{BH}}^2)^{\frac{s+1}{2}}R_{\ell m}(r),\\
a_0&=i\,\frac{\omega -m \Omega_H}{4 \pi  T_H}-2 i M \omega -\frac{s}{2},\\
a_1&=i\,\frac{\omega -m \Omega_H}{4 \pi  T_H}+\frac{s}{2},\\
\epsilon&=-16\, i\, \pi\, M\,\omega\, R_h\, T_H ,\\
\mu&=s-2i\,M\,\omega,\\
u&=-{}_sA_{\ell\,m}-s(s+1) +7 M^2 \omega ^2\\
&\ +8\, \pi\,M\, \omega  \,R_h\, T_H   (2 M\, \omega +i s)+(4 \pi\, M \,\omega\, R_h\, T_H)^2,
\end{aligned}
\end{equation}
where ${}_sA_{\ell\,m}$ is the separation constant, whose quantization can be obtained by solving the associated angular problem and imposing the regularity conditions at $\theta=0$ and $\theta=\pi$.
For example, the expansion of the separation constant around $a_{\text{BH}}\,\omega=0$ reads \footnote{This is a convergent series \cite{Arnaudo:2022ivo} whose coefficients can be computed explicitly, see \cite{Aminov:2020yma,Bonelli:2021uvf} for details.}
\begin{equation}\label{separationconstant}
\begin{aligned}
&{}_sA_{\ell\,m}=(\ell-s) (\ell+s+1)-\frac{2 m\, s^2\, \omega\,a_{\text{BH}} }{\ell(\ell+1)}
+\mathcal{O}\left(a_{\text{BH}}^2\omega^2\right).
\end{aligned}
\end{equation}
Notice that the angular problem does not depend at all on $M$, and so the expression of the separation constant is not affected by passing to the (near-)extremal limit.
Moreover, we find it useful to introduce the following quantities, which are linear combinations of the parameters $a_0$ and $a_1$:\footnote{These quantities $\mu_1$ and $\mu_2$, together with the parameter $\mu$, are the masses of the hypermultiplets in the fundamental representation from the gauge theory point of view.}
\begin{equation}\label{twomassesgauge}
\begin{aligned}
\mu_1=&\,a_0-a_1=-s-2i\,M\,\omega,\\
\mu_2=&\,-a_0-a_1=2 i\, M\, \omega -i\,\frac{\omega -m \Omega_H}{2 \pi  T_H}.
\end{aligned}
\end{equation}


Here, we consider the relevant connection problem for the radial problem to obtain the quantization condition determining the QNMs.
It is useful to rewrite the above equation with the redefinition
\begin{equation}
w(z)=e^{-\epsilon z/2} z^{-\gamma/2} (1-z)^{-\delta/2}\psi(z),
\end{equation}
so that the confluent Heun differential equation becomes
\begin{equation}
\frac{d^2w}{dz^2}+\left(\frac{\gamma}{z}+\frac{\delta}{z-1}+\epsilon\right)\frac{dw}{dz}+\frac{\alpha z-q}{z(z-1)}w=0\,,
\end{equation}
with
\begin{equation}
\begin{aligned}
    & \gamma = 1 - 2 a_0 \,, \\
    & \delta = 1 - 2 a_1\,, \\
    & \alpha = \mu\,\epsilon+\frac{\gamma+\delta}{2}\,\epsilon \,, \\
    & q = \frac{1}{4}-u+\alpha - \frac{(\gamma+\delta-1)^2}{4}-\frac{\delta\, \epsilon}{2}\,.
\end{aligned}
\end{equation}
The boundary conditions require the presence of only ingoing modes at the black hole horizon and only outgoing modes at radial infinity. In terms of the function $w$, these translate into the following requirements:
\begin{equation}
\begin{aligned}
&w(z)\sim 1\ \ \ \ \ \ \ \ \ \ \ \ \ \quad\text{for}\ \ z\sim 1,\\
&w(z)\sim \mathrm{e}^{-\epsilon z} z^{\frac{\alpha}{\epsilon}-\gamma-\delta}\ \ \text{for}\ \ z\sim \infty.
\end{aligned}
\end{equation}
The connection formula from the irregular singularity at $z=\infty$ to the regular singularity at $z=1$ reads 
\cite{Bonelli:2022ten}:
\begin{equation}\label{connectioninftyto1}
\begin{aligned}
&e^{-\epsilon z} z^{\frac{\alpha}{\epsilon}-\gamma-\delta} \mathrm{HeunC}_\infty(q-\gamma\epsilon,\alpha-\epsilon(\gamma+\delta),\gamma,\delta,-\epsilon;z)=\\
&\epsilon^{\frac{1}{2}-\mu}e^{-\frac{1}{2}\partial_{\mu}F}\times\\
&\biggl[\sum_{\sigma=\pm}\frac{\Gamma\brc{-2\sigma a}\Gamma\brc{1-2\sigma\,a}\Gamma\brc{2a_1}\epsilon^{\sigma a}e^{-\frac{\sigma}{2}\partial_{a}F+\frac{1}{2}\partial_{a_1}F}}{\Gamma\brc{\frac{1}{2}-\mu_1-\sigma a}\Gamma\brc{\frac{1}{2}-\mu_2-\sigma a}\Gamma\brc{\frac{1}{2}-\mu-\sigma a}}\\
&\times\mathrm{HeunC}(q-\alpha,-\alpha,\delta,\gamma,-\epsilon;1-z) \\
&+\sum_{\sigma=\pm}\frac{\Gamma\brc{-2\sigma a}\Gamma\brc{1-2\sigma\,a}\Gamma\brc{-2a_1}\epsilon^{\sigma a}e^{-\frac{\sigma}{2}\partial_{a}F-\frac{1}{2}\partial_{a_1}F}}{\Gamma\brc{\frac{1}{2}+\mu_1-\sigma a}\Gamma\brc{\frac{1}{2}+\mu_2-\sigma a}\Gamma\brc{\frac{1}{2}-\mu-\sigma a}}\\
&\times  (1-z)^{1-\delta} \mathrm{HeunC}\left(\tilde{q},-\alpha-(1-\delta)\epsilon, 2-\delta,\gamma,-\epsilon;1-z \right)\biggr],
\end{aligned}
\end{equation}
where $\tilde{q}=q-\alpha-(1-\delta)(\epsilon+\gamma)$ and $a$ is the composite monodromy parameter around the singularities at $z=0$ and $z=1$. We refer to the Appendix for the relevant notations.

As computed in \eqref{detresult}, the contribution of QNMs to the determinant \footnote{Strictly speaking the GY formula computes the regularised determinant  without zero modes ${\rm det}'$. For $s=1,2$ the operator we consider do not display zero modes so that 
${\rm det}'={\rm det}$.} at fixed quantum numbers $\ell,m,s$ is
\begin{equation}\label{detQNMs}
\begin{aligned}
&\mathrm{det}_{(\ell,m,s)}^{\text{[Kerr]}}=\sqrt{2\,\pi}\\
&\times\sum_{\sigma=\pm}\frac{\Gamma\brc{-2\sigma a}\Gamma\brc{1-2\sigma a}\epsilon^{\frac{1}{2}-\mu+\sigma a}e^{-\frac{1}{2}\left(\sigma\partial_a+\partial_{a_1}+\partial_{\mu}\right)F}}{\Gamma\brc{\frac{1}{2}+\mu_1-\sigma a}\Gamma\brc{\frac{1}{2}+\mu_2-\sigma a}\Gamma\brc{\frac{1}{2}-\mu-\sigma a}}.
\end{aligned}
\end{equation}
Let us remark that the above formula displays a branch cut in $\omega$ due to the presence of the term $\epsilon^{\frac{1}{2}-\mu+\sigma a}$, and the dictionary \eqref{dictionarykerr}.

The exact quantization condition for QNMs follows by studying the zeros of \eqref{detQNMs}. This corresponds to
\begin{equation}\label{newquantcond}
\begin{aligned}
&\frac{\Gamma(1-2a)^2\Gamma\left(\frac{1}{2}+\mu_1+a\right)\Gamma\left(\frac{1}{2}-\mu+a\right)}{\Gamma(1+2a)^2\Gamma\left(\frac{1}{2}+\mu_1-a\right)\Gamma\left(\frac{1}{2}-\mu-a\right)}e^{-\partial_aF}\\
&\times\frac{\Gamma\left(\frac{1}{2}+\mu_2+a\right)}{\Gamma\left(\frac{1}{2}+\mu_2-a\right)}\epsilon^{2a}=1.
\end{aligned}
\end{equation}
In the confluent limit corresponding to the extremal Kerr case in which the horizons $R_h$ and $R_i$ coalesce,  $\epsilon\to 0$, and $\mu_2\to\infty$, in such a way that $\Lambda\equiv\epsilon\,\mu_2$ is finite 
\footnote{From the gauge theory viewpoint, this procedure corresponds to the holomorphic decoupling of the hypermultiplet mass $\mu_2$, and it produces the $N_f=2$ theory from the original $N_f=3$ one. The new parameter $\Lambda\equiv -8 M^2 \omega  (\omega-m\,\Omega_H^{\text{ext}})$, with $\Omega_H^{\text{ext}}=1/(2M)$, corresponds to the instanton counting parameter of the $N_f=2$ theory.}. In this limit, 
\begin{equation}\label{decoupling}
\begin{aligned}
&\frac{\Gamma\left(\frac{1}{2}+\mu_2+a\right)}{\Gamma\left(\frac{1}{2}+\mu_2-a\right)}\epsilon^{2a}\sim\left[\left(\frac{1}{2}+\mu_2-a\right)\epsilon\right]^{2a}= \Lambda^{2a}+\mathcal{O}(\epsilon),
\end{aligned}
\end{equation}
where 
we used the Stirling approximation $\Gamma(z+\alpha)/\Gamma(z)\sim z^{\alpha}$ as $z\to\infty$ and $\alpha\in\mathbb{C}$.
The expansion of the parameter $a$ is determined up to an overall sign and reads
\begin{equation}\label{zeroinsta}
\begin{aligned}
&a=\pm\sqrt{\frac{1}{4}-u}+\mathcal{O}(\epsilon).
\end{aligned}
\end{equation}
The choice of the branch in \eqref{zeroinsta} is a symmetry of the theory, and we choose the positive one to ensure that $\epsilon^{a}\to 0$ in the small $\epsilon$ expansion.

As we see from \eqref{decoupling}, in the extremal limit an entire branch of 
solutions associated to the poles of the $\Gamma$ function in the second line of \eqref{newquantcond}
decouples from the spectrum.
From \eqref{detQNMs}, it is clear that one of the two channels of the connection problem gets suppressed, and \eqref{newquantcond}, up to corrections of order $\epsilon^{2a}$, is solved by the poles of the $\Gamma$-function in the remaining channel, which is precisely the one appearing in the numerator of \eqref{decoupling}:
\begin{equation}\label{zdmquantcond}
\frac{1}{2}+a+\mu_2=-n,\quad n\in\mathbb{Z}_{\ge 0}.
\end{equation}

The physical properties of this branch of modes become clear by using the dictionary \eqref{dictionarykerr}
to the gravitational quantities. Let us consider the following small temperature expansions 
\begin{equation}\label{smallTexpansion}
\begin{aligned}
a&=\sum_{j\ge 0}a^{(j)}\,T_H^j,\\
{}_sA_{\ell m}&=\sum_{j\ge 0}{}_sA_{\ell m}^{(j)}\,T_H^j,\\
\omega^*_{n,\ell,m}&=\sum_{j\ge 0}\frac{\beta_j}{a_{\text{BH}}}\,T_H^j,
\end{aligned}
\end{equation}
where we denoted with $\omega^*_{n,\ell,m}$ the series expansion in the temperature of the frequency solutions to \eqref{zdmquantcond}.
Since $\mu_2$ diverges for $a_{\text{BH}}\to M$, the leading order of the lhs of \eqref{zdmquantcond} is finite if (see \eqref{twomassesgauge})
\begin{equation}\label{omega0}
\beta_0=m/2.
\end{equation}
Then, inserting \eqref{smallTexpansion} and \eqref{omega0} in \eqref{zeroinsta}, we find
\begin{equation}
\begin{aligned}
a^{(0)}&=\frac{1}{2}\sqrt{4{}_sA_{\ell\,m}^{(0)}-7m^2+(1+2s)^2},
\end{aligned}
\end{equation}
where
\begin{equation}
{}_sA_{\ell m}^{(0)}
={}_sA_{\ell m}\bigg|_{\omega=m/(2a_{\text{BH}})}.
\end{equation}
With the substitutions \eqref{smallTexpansion} and \eqref{omega0} in \eqref{twomassesgauge}, we find
\begin{equation}
\mu_2=-i\sqrt{2M}\beta_1+\mathcal{O}\left(T_H\right).
\end{equation}
Hence, the quantization condition \eqref{zdmquantcond} gives
at leading order
\begin{equation}\label{omega1}
\beta_1=-i\,\frac{2n+1+\sqrt{4{}_sA_{\ell\,m}^{(0)}-7m^2+(1+2s)^2}}{2\sqrt{2M}}.
\end{equation}
Substituting \eqref{omega0} and \eqref{omega1} in \eqref{smallTexpansion}, we get
\begin{equation}\label{omegaresult}
\begin{aligned}
&\omega^*_{n,\ell,m}=\frac{m}{2a_{\text{BH}}}\\
&-i\,\pi\,T_H\brc{2n+1+\sqrt{4{}_sA_{\ell\,m}^{(0)}-7m^2+(1+2s)^2}}\\
&+\mathcal{O}\brc{T_H^2},
\end{aligned}
\end{equation}
which has a parametrically small negative imaginary part, in agreement with formula (59) of \cite{Cavalcante:2024kmy}, proportional to the temperature $T_H$.
These modes have a small imaginary part in the near-extremal limit and are called
zero-damping modes (ZDMs).
Our analysis is consistent with the WKB computations presented in \cite{Yang:2013uba} and \cite{Casals:2019vdb}, and the suppression $\epsilon^{2a}\to 0$ is in accordance with the results in \cite{Berti:2005gp}.

A particular sub-branch of modes can be obtained for $m=0$. In this case, the expansion of the separation constant ${}_sA_{\ell m}$ around $a_{\text{BH}}\,\omega=0$ in \eqref{separationconstant} becomes relevant. Indeed, in the near-extremal limit
\begin{equation}
a_{\text{BH}}\omega^*_{n,\ell,m}=\frac{m}{2}+\mathcal{O}\left(T_H\right),
\end{equation}
and this scales with the temperature for $m=0$. It follows that \eqref{omegaresult} simplifies to
\begin{equation}\label{Hod}
\omega^*_{n,\ell,0}=-i\brc{n+\ell+1}2\,\pi\,T_H+\mathcal{O}\brc{T_H^2},
\end{equation}
which were considered also in \cite{Hod:2013fea}.

Let us underline that the $\Gamma$-function that gives the modes \eqref{omegaresult} arises in the step of the connection procedure which gets performed locally around the black hole horizon (for the full details 
about the confluent Heun connection problem  we refer to formulae (3.2.11) and (3.2.12) in \cite{Bonelli:2022ten}).
Therefore, our result does not depend on the near-infinity region and applies to any near-extremal regime of black hole backgrounds independently from their asymptotic geometry.
This proves the universality of the appearance of ZDMs in near-extremal geometries in the near-horizon region.

Let us now focus on the contribution to the effective action of the ZDMs and compute the leading contribution in the Hawking temperature they imply.

Let us start by studying the normalization of \eqref{detQNMs}.
From \eqref{decoupling}, we have
\begin{equation}\label{prefactorization}
\frac{\epsilon^{-a}\,\Lambda^{a}}{\Gamma\left(\frac{1}{2}+\mu_2+a\right)}\sim \frac{\epsilon^{a}\,\Lambda^{-a}}{\Gamma\left(\frac{1}{2}+\mu_2-a\right)}\left[1-\Lambda^{-2a}\mathcal{O}(\epsilon)\right],
\end{equation}
and this permits us to rewrite \eqref{detQNMs} in the factorized form
\begin{equation}\label{detfactorized}
\mathrm{det}_{(\ell,m,s)}^{[\text{NEK}]}=\mathrm{det}_{(\ell,m,s)}^{[\text{ZDM}]}\left[\mathrm{det}_{(\ell,m,s)}^{[\text{EK}]}+\mathcal{O}\left(\epsilon\right)\right],
\end{equation}
where
\begin{align}
\mathrm{det}_{(\ell,m,s)}^{[\text{ZDM}]}&=\frac{\sqrt{2\,\pi}\,\epsilon^{\frac{1}{2}-\mu}\,\left(\epsilon/\Lambda\right)^{a}}{\Gamma\left(\frac{1}{2}+\mu_2-a\right)},\label{detZDMfact}\\
\mathrm{det}_{(\ell,m,s)}^{[\text{EK}]}&=\sum_{\sigma=\pm}\frac{\Gamma\brc{-2\sigma a}\Gamma\brc{1-2\sigma a}\Lambda^{\sigma a}e^{-\frac{1}{2}\left(\sigma \partial_a+\partial_{a_1}+\partial_{\mu}\right)F}}{\Gamma\brc{\frac{1}{2}+\mu_1-\sigma a}\Gamma\brc{\frac{1}{2}-\mu-\sigma a}}
.\label{detExtfact}
\end{align}
 Eq.\eqref{detZDMfact}
is the contribution of the ZDMs which are getting decoupled in the extremal limit, and
\eqref{detExtfact} is the QNM-determinant of the doubly-confluent Heun equation, encoding the spectral problem of the extremal Kerr case. Thanks to \eqref{prefactorization}, the definition of the normalization \eqref{detZDMfact} is independent of the branch of \eqref{zeroinsta} up to orders in $\epsilon$. In \eqref{detZDMfact}, for definiteness, we chose the positive branch. The terms in \eqref{detZDMfact} characterize the corrections of the near-extremal geometry, including the normalization factors of the confluent Heun solutions.
Indeed,
when considering the extremal Kerr limit one must redefine the $z$ variable. As can be seen from \eqref{dictionarykerr}, in this limit, $z\to\infty$ and $\epsilon\to 0$ with the local new variable $\epsilon\,z$ finite. The redefinition of the local variable implies a Jacobian factor that cancels the term $\epsilon^{\frac{1}{2}-\mu+a}$ in the numerator of \eqref{detZDMfact} \footnote{In detail, this extra factor can be seen as the combination of an overall factor $\epsilon^{\frac{1}{2}-\mu}$ in the normalization of the irregular semiclassical block around infinity and a factor $\epsilon^a$ coming from the irregular semiclassical block around zero with shifted intermediate momentum, arising in the confluence procedure between the singularities at $z=0$ and $z=1$. Specifically, the first factor can be seen by taking the semiclassical limit of the $\Lambda$-factors in the first line of formula (3.2.7) in \cite{Bonelli:2022ten}. In particular, considering the semiclassical limit $b\to 0$, the term $b\,\Lambda$ corresponds to our parameter $\epsilon$, and the semiclassical limit of
$\Delta_{2,1}-\theta\,b\,\mu$ in \cite{Bonelli:2022ten} becomes $-\frac{1}{2}+\mu$ in our notation. This, consistently with our choices of the local solution around $z=\infty$, cancels the factor $\epsilon^{\frac{1}{2}-\mu}$. The other factor can be seen by taking the semiclassical limit of the $\Lambda_2$-term in formula (3.4.4) in \cite{Bonelli:2022ten}, when considering the shift of the intermediate momentum. Indeed, $\Delta_{\sigma}\to \sigma\,a$ in the semiclassical limit, and this simplifies with the corresponding factor in the connection formula \eqref{connectioninftyto1}. This, in turn, implies the simplification of the term $\epsilon^a$ in \eqref{detfactorized} after taking the confluent limit.}. The gauge quantities $a$ and $F$ in \eqref{detExtfact} are redefined as the corresponding quantities in the $N_f=2$ gauge theory. In \eqref{detExtfact} the expressions for $\mu$ and $\mu_1$ remain the same as appearing in \eqref{dictionarykerr} and \eqref{twomassesgauge}, and $\Lambda=-8M^2\omega\left(\omega-m\Omega_H^{\text{ext}}\right)$, with $\Omega_H^{\text{ext}}=1/(2M)$.

To determine the temperature contribution to the partition function, we therefore focus on the $\Gamma$-function in \eqref{detZDMfact}, and, inspired by the DHS formula \cite{Denef:2009kn} and its generalization to non-static spacetimes \cite{Castro:2017mfj}, we substitute the gravitational dictionary \eqref{dictionarykerr} evaluated in the Matsubara frequencies. Moreover, we focus on the case $s>0$, specifying in the end how the analysis applies to the case $s<0$.
By imposing the exponent of the local solution in \eqref{local1} to be a nonnegative integer, we find the following quantization for the Matsubara frequencies associated with the QNMs:
\begin{equation}
2a_1=-k,\quad k\ge 0,
\end{equation}
from which, using \eqref{dictionarykerr},
\begin{equation}\label{MatsubaraKerr}
\omega_k^{(M)}=m\Omega_H+2\pi i T_H (k+s).
\end{equation}
This selects the frequencies with good Euclidean continuation in the Kerr geometry, and is analogous to the results found in \cite{Castro:2017mfj} for $s=2$ perturbations around the rotating BTZ black hole.

By using the gravitational dictionary \eqref{dictionarykerr}, the expression of the Matsubara frequencies \eqref{MatsubaraKerr}, and the first-instanton expansion for $a$:
\begin{equation}
\begin{aligned}
a=\sqrt{\frac{1}{4}-u}-\frac{u+a_0^2-a_1^2}{4u\sqrt{\frac{1}{4}-u}}\mu\,\epsilon+\mathcal{O}\left(\epsilon^2\right),
\end{aligned}
\end{equation}
we find
\begin{equation}
\mu_2\bigg|_{\omega=\omega_k^{(M)}}=k+s+2i\,M\left[m \Omega_H+2\pi i\,T_H\left(k+s\right)\right]
\end{equation}
so that
\begin{equation}\label{eqwithc1m}
\begin{aligned}
&\frac{1}{2}+\mu_2-a\bigg|_{\omega=\omega_k^{(M)}}=\frac{1}{2}+k+s+i\,m\\
&-\frac{1}{2} \sqrt{4 {}_sA_{\ell m}^{(0)}-7 m^2+(2 s+1)^2}+c_1(m)\,T_H+\mathcal{O}(T_H^2),
\end{aligned}
\end{equation}
where $c_1(m)$ is given in \eqref{c1m}. When $m=0$, by comparison with the expansion for the separation constant ${}_sA_{\ell 0}$ around $a_{\text{BH}}\,\omega=0$ in \eqref{separationconstant}, we obtain
\begin{equation}\label{c10}
c_1(0)=-4\pi\, M \left(k+s\right),
\end{equation}
where we used that, in the near-extremal limit,
\begin{equation}
a_{\text{BH}}\,\omega^{(M)}_k=\mathcal{O}\left(T_H\right)
\end{equation}
for $m=0$. 
We remark that for $m=0$ and $\omega=\omega_k^{(M)}$ the parameter $\epsilon\sim\mathcal{O}\left(T_H^2\right)$ in the small $T_H$-expansion.
We finally get
\begin{equation}
\frac{1}{2}+\mu_2-a\,\bigg|_{\omega=\omega_k^{(M)},\, m=0}=k+s-\ell-4\pi M T_H (k+s)+\mathcal{O}\left(T_H^2\right).
\end{equation}

Taking the product over the quantum numbers $k, \ell, m$, we thus find that the inverse of the determinant coming from \eqref{detZDMfact}, up to factors independent of the temperature, is equal to
\begin{widetext}
\begin{equation}\label{temperaturedependence}
\begin{aligned}
&\prod_{\ell\ge s}\prod_{m=-\ell}^{\ell}\prod_{k\ge 0}\left[\frac{1}{\mathrm{det}^{[\text{ZDM}]}_{(\ell,m,s)}}\bigg|_{\omega=\omega_k^{(M)}}\right]\sim\prod_{\ell\ge s}\prod_{m=-\ell}^{\ell}\prod_{k\ge 0}\Gamma\left(\frac{1}{2}+\mu_2-a\right)\bigg|_{\omega=\omega_k^{(M)}}=\\
&\prod_{\ell\ge s}\prod_{m=-\ell}^{\ell}\prod_{k\ge 0}\Gamma\left(\frac{1}{2}+k+s+i\,m-\frac{1}{2} \sqrt{4 {}_sA_{\ell m}^{(0)}-7 m^2+(2 s+1)^2}+\mathcal{O}(T_H)\right)=\\
&\prod_{\ell\ge s}\prod_{0\ne m=-\ell}^{\ell}\prod_{k\ge 0}\Gamma\left(\frac{1}{2}+k+s+i\,m-\frac{1}{2} \sqrt{4 {}_sA_{\ell m}^{(0)}-7 m^2+(2 s+1)^2}+\mathcal{O}(T_H)\right)\times\\
&\prod_{\ell\ge s}\ \prod_{k\ge 0}\Gamma\left(k+s-\ell-4\pi M T_H (k+s)+\mathcal{O}\left(T_H^2\right)\right),
\end{aligned}
\end{equation}
\end{widetext}
where in the second equality we split the contributions of $m\ne 0$ and $m=0$.
Eq.\eqref{temperaturedependence} gives small contributions in the low-temperature limit only for the poles of the $\Gamma$-functions.
For this reason, the factors corresponding to $m\ne 0$ give finite contributions in the small temperature limit, their argument being different from nonpositive integers. Indeed, ${}_sA_{\ell m}^{(0)}$ evaluated in the leading order in $T_H$ of \eqref{MatsubaraKerr}, that is $m\,\Omega_H$, is real, and so is the term under the square root. Then, if the square root is real, in the argument of the $\Gamma$-function there is an imaginary part given by $i\,m$, whereas if the square root is imaginary, the argument of the $\Gamma$-function has a positive real part $\frac{1}{2}+k+s$.
Therefore the leading factors in the low-temperature behaviour come from 
the last line. By neglecting subleading corrections in the temperature, we have
\begin{widetext}
\begin{equation}\label{infiniteproducts}
\begin{aligned}
&\prod_{\ell\ge s}\ \prod_{k\ge 0}\Gamma\left(k+s-\ell-4\pi M T_H (k+s)\right)=\prod_{\ell'\ge 0}\ \prod_{k\ge 0}\Gamma\left(k-\ell'-4\pi M T_H (k+s)\right)=\\
&\prod_{k\ge 0}\ \prod_{r\in\mathbb{Z}}\Gamma\left(r-4\pi M T_H (k+s)\right)\sim \prod_{k\ge 0}\ \prod_{r\in\mathbb{Z}}\ \prod_{n\ge 0}\left(n+r-4\pi M T_H (k+s)\right)^{-1}=\\
&\prod_{r\in\mathbb{Z}}\prod_{ n+r\ne 0}\prod_{k\ge 0}\left(n+r-4\pi M T_H (k+s)\right)^{-1}\prod_{n\ge 0}\prod_{k\ge 0}\left(-4\pi M T_H (k+s)\right)^{-1}.
\end{aligned}
\end{equation}
\end{widetext}
In the above computation, we first introduced $\ell'=\ell-s$, then $r=k-\ell'$, and we used the infinite product representation of the $\Gamma$-function. 
Finally, we can select the last infinite product in \eqref{infiniteproducts} corresponding to $r=-n$ as the only one scaling with the temperature, and rewrite it as
\begin{equation}\label{secondstepinfiniteproducts}
\begin{aligned}
&\prod_{n\ge 0}\prod_{k\ge 0}\left(-4\pi M T_H (k+s)\right)^{-1}\sim\\
&\prod_{n\ge 0}\Gamma_1\left(4\pi\,s\,M\,T_H\mid 4\pi\,M\,T_H\right)=\\
&\prod_{n\ge 0}\frac{\Gamma(s)}{\sqrt{2\pi}}\left(4\pi\,M\,T_H\right)^{s-\frac{1}{2}}\sim
T_H^{\frac{s}{2}-\frac{1}{4}}
\end{aligned}
\end{equation}
where we used
\begin{equation}
\Gamma_1(z|a)=\frac{a^{a^{-1}z-\frac{1}{2}}}{\sqrt{2\pi}}\Gamma(a^{-1}z)
\end{equation}
and in the last line we extracted the explicit $T_H$-dependence by a further $\zeta$-regularization
$\prod_{n\ge 0}T_H=T_H^{1/2}$.



The gravitational perturbations are described by the Teukolsky equation with 
 $s=\pm 2$ which correspond to the two helicity states of the on-shell graviton. They give both the same contributions thanks to the symmetry of the separation constant
\begin{equation}
{}_{-s}A_{\ell m}={}_{s}A_{\ell m}+2s
\end{equation}
and the fact that $s\to -s$ swaps QNMs and anti-QNMs in the analysis of Matsubara frequencies \footnote{The Matsubara frequencies associated with the anti-QNMs are obtained by taking the asymptotic behaviour $\psi_+^{(1)}$ in \eqref{localsolutionsat1} which, together with the corresponding normalization, amounts to impose $2a_1=k$, $k\ge 0$, so that $\omega_k^{(M)}(-s)=\overline{\omega}_k^{(M)}(s)$.}.
Taking into account that the one-loop partition function is proportional to the inverse square root of the determinant, we conclude 
\begin{equation}
Z^{\text{NEK}}_{\text{1-loop}}\sim \frac{1}{\sqrt{|\mathrm{det}^{[\text{ZDM}]}|^2}}\frac{1}{\sqrt{|\mathrm{det}^{[\text{EK}]}|^2}}\sim T_H^{\frac{3}{2}} \,Z^{\text{EK}}_{\text{1-loop}},   
\end{equation}
where $Z^{\text{EK}}_{\text{1-loop}}$ is the one-loop partition function of the extremal Kerr geometry.
Analogously, the temperature scaling we find for $s=1$ is $T_H^{\frac{1}{2}}$.

\paragraph{\bf Acknowledgments:}
We would like to thank 
Davide Cassani, 
Cristoforo Iossa,
Leopoldo Pando Zayas,
Mukund Rangamani
and
Chiara Toldo 
for useful discussions,  clarifications and/or a careful reading of the manuscript.
        We would like to thank also the organizers of the conference "TFI 2024: Theories of the Fundamental Interactions" (Napoli 23–25 Sept 2024)
        for the fertile scientific atmosphere which inspired this letter.
		This research  is  partially supported by the INFN Research Projects GAST and ST$\&$FI, and  by PRIN
		"String theory as a bridge between Gauge Theories and Quantum Gravity".

		\bibliography{biblio}

\begin{thebibliography}{57}%
\makeatletter
\providecommand \@ifxundefined [1]{%
 \@ifx{#1\undefined}
}%
\providecommand \@ifnum [1]{%
 \ifnum #1\expandafter \@firstoftwo
 \else \expandafter \@secondoftwo
 \fi
}%
\providecommand \@ifx [1]{%
 \ifx #1\expandafter \@firstoftwo
 \else \expandafter \@secondoftwo
 \fi
}%
\providecommand \natexlab [1]{#1}%
\providecommand \enquote  [1]{``#1''}%
\providecommand \bibnamefont  [1]{#1}%
\providecommand \bibfnamefont [1]{#1}%
\providecommand \citenamefont [1]{#1}%
\providecommand \href@noop [0]{\@secondoftwo}%
\providecommand \href [0]{\begingroup \@sanitize@url \@href}%
\providecommand \@href[1]{\@@startlink{#1}\@@href}%
\providecommand \@@href[1]{\endgroup#1\@@endlink}%
\providecommand \@sanitize@url [0]{\catcode `\\12\catcode `\$12\catcode `\&12\catcode `\#12\catcode `\^12\catcode `\_12\catcode `\%12\relax}%
\providecommand \@@startlink[1]{}%
\providecommand \@@endlink[0]{}%
\providecommand \url  [0]{\begingroup\@sanitize@url \@url }%
\providecommand \@url [1]{\endgroup\@href {#1}{\urlprefix }}%
\providecommand \urlprefix  [0]{URL }%
\providecommand \Eprint [0]{\href }%
\providecommand \doibase [0]{http://dx.doi.org/}%
\providecommand \selectlanguage [0]{\@gobble}%
\providecommand \bibinfo  [0]{\@secondoftwo}%
\providecommand \bibfield  [0]{\@secondoftwo}%
\providecommand \translation [1]{[#1]}%
\providecommand \BibitemOpen [0]{}%
\providecommand \bibitemStop [0]{}%
\providecommand \bibitemNoStop [0]{.\EOS\space}%
\providecommand \EOS [0]{\spacefactor3000\relax}%
\providecommand \BibitemShut  [1]{\csname bibitem#1\endcsname}%
\let\auto@bib@innerbib\@empty
\bibitem [{\citenamefont {Preskill}\ \emph {et~al.}(1991)\citenamefont {Preskill}, \citenamefont {Schwarz}, \citenamefont {Shapere}, \citenamefont {Trivedi},\ and\ \citenamefont {Wilczek}}]{Preskill:1991tb}%
  \BibitemOpen
  \bibfield  {author} {\bibinfo {author} {\bibfnamefont {J.}~\bibnamefont {Preskill}}, \bibinfo {author} {\bibfnamefont {P.}~\bibnamefont {Schwarz}}, \bibinfo {author} {\bibfnamefont {A.~D.}\ \bibnamefont {Shapere}}, \bibinfo {author} {\bibfnamefont {S.}~\bibnamefont {Trivedi}}, \ and\ \bibinfo {author} {\bibfnamefont {F.}~\bibnamefont {Wilczek}},\ }\href {\doibase 10.1142/S0217732391002773} {\bibfield  {journal} {\bibinfo  {journal} {Mod. Phys. Lett. A}\ }\textbf {\bibinfo {volume} {6}},\ \bibinfo {pages} {2353} (\bibinfo {year} {1991})}\BibitemShut {NoStop}%
\bibitem [{\citenamefont {Banerjee}\ \emph {et~al.}(2011{\natexlab{a}})\citenamefont {Banerjee}, \citenamefont {Gupta},\ and\ \citenamefont {Sen}}]{Banerjee:2010qc}%
  \BibitemOpen
  \bibfield  {author} {\bibinfo {author} {\bibfnamefont {S.}~\bibnamefont {Banerjee}}, \bibinfo {author} {\bibfnamefont {R.~K.}\ \bibnamefont {Gupta}}, \ and\ \bibinfo {author} {\bibfnamefont {A.}~\bibnamefont {Sen}},\ }\href {\doibase 10.1007/JHEP03(2011)147} {\bibfield  {journal} {\bibinfo  {journal} {JHEP}\ }\textbf {\bibinfo {volume} {03}},\ \bibinfo {pages} {147} (\bibinfo {year} {2011}{\natexlab{a}})},\ \Eprint {http://arxiv.org/abs/1005.3044} {arXiv:1005.3044 [hep-th]} \BibitemShut {NoStop}%
\bibitem [{\citenamefont {Banerjee}\ \emph {et~al.}(2011{\natexlab{b}})\citenamefont {Banerjee}, \citenamefont {Gupta}, \citenamefont {Mandal},\ and\ \citenamefont {Sen}}]{Banerjee:2011jp}%
  \BibitemOpen
  \bibfield  {author} {\bibinfo {author} {\bibfnamefont {S.}~\bibnamefont {Banerjee}}, \bibinfo {author} {\bibfnamefont {R.~K.}\ \bibnamefont {Gupta}}, \bibinfo {author} {\bibfnamefont {I.}~\bibnamefont {Mandal}}, \ and\ \bibinfo {author} {\bibfnamefont {A.}~\bibnamefont {Sen}},\ }\href {\doibase 10.1007/JHEP11(2011)143} {\bibfield  {journal} {\bibinfo  {journal} {JHEP}\ }\textbf {\bibinfo {volume} {11}},\ \bibinfo {pages} {143} (\bibinfo {year} {2011}{\natexlab{b}})},\ \Eprint {http://arxiv.org/abs/1106.0080} {arXiv:1106.0080 [hep-th]} \BibitemShut {NoStop}%
\bibitem [{\citenamefont {Sen}(2012{\natexlab{a}})}]{Sen:2012kpz}%
  \BibitemOpen
  \bibfield  {author} {\bibinfo {author} {\bibfnamefont {A.}~\bibnamefont {Sen}},\ }\href {\doibase 10.1007/s10714-012-1336-5} {\bibfield  {journal} {\bibinfo  {journal} {Gen. Rel. Grav.}\ }\textbf {\bibinfo {volume} {44}},\ \bibinfo {pages} {1207} (\bibinfo {year} {2012}{\natexlab{a}})},\ \Eprint {http://arxiv.org/abs/1108.3842} {arXiv:1108.3842 [hep-th]} \BibitemShut {NoStop}%
\bibitem [{\citenamefont {Sen}(2012{\natexlab{b}})}]{Sen:2012cj}%
  \BibitemOpen
  \bibfield  {author} {\bibinfo {author} {\bibfnamefont {A.}~\bibnamefont {Sen}},\ }\href {\doibase 10.1007/s10714-012-1373-0} {\bibfield  {journal} {\bibinfo  {journal} {Gen. Rel. Grav.}\ }\textbf {\bibinfo {volume} {44}},\ \bibinfo {pages} {1947} (\bibinfo {year} {2012}{\natexlab{b}})},\ \Eprint {http://arxiv.org/abs/1109.3706} {arXiv:1109.3706 [hep-th]} \BibitemShut {NoStop}%
\bibitem [{\citenamefont {Bhattacharyya}\ \emph {et~al.}(2014)\citenamefont {Bhattacharyya}, \citenamefont {Grassi}, \citenamefont {Marino},\ and\ \citenamefont {Sen}}]{Bhattacharyya:2012ye}%
  \BibitemOpen
  \bibfield  {author} {\bibinfo {author} {\bibfnamefont {S.}~\bibnamefont {Bhattacharyya}}, \bibinfo {author} {\bibfnamefont {A.}~\bibnamefont {Grassi}}, \bibinfo {author} {\bibfnamefont {M.}~\bibnamefont {Marino}}, \ and\ \bibinfo {author} {\bibfnamefont {A.}~\bibnamefont {Sen}},\ }\href {\doibase 10.1088/0264-9381/31/1/015012} {\bibfield  {journal} {\bibinfo  {journal} {Class. Quant. Grav.}\ }\textbf {\bibinfo {volume} {31}},\ \bibinfo {pages} {015012} (\bibinfo {year} {2014})},\ \Eprint {http://arxiv.org/abs/1210.6057} {arXiv:1210.6057 [hep-th]} \BibitemShut {NoStop}%
\bibitem [{\citenamefont {Pando~Zayas}\ and\ \citenamefont {Xin}(2019)}]{PandoZayas:2019hdb}%
  \BibitemOpen
  \bibfield  {author} {\bibinfo {author} {\bibfnamefont {L.~A.}\ \bibnamefont {Pando~Zayas}}\ and\ \bibinfo {author} {\bibfnamefont {Y.}~\bibnamefont {Xin}},\ }\href {\doibase 10.1103/PhysRevD.100.126019} {\bibfield  {journal} {\bibinfo  {journal} {Phys. Rev. D}\ }\textbf {\bibinfo {volume} {100}},\ \bibinfo {pages} {126019} (\bibinfo {year} {2019})},\ \Eprint {http://arxiv.org/abs/1908.01194} {arXiv:1908.01194 [hep-th]} \BibitemShut {NoStop}%
\bibitem [{\citenamefont {Benini}\ \emph {et~al.}(2020)\citenamefont {Benini}, \citenamefont {Gang},\ and\ \citenamefont {Pando~Zayas}}]{Benini:2019dyp}%
  \BibitemOpen
  \bibfield  {author} {\bibinfo {author} {\bibfnamefont {F.}~\bibnamefont {Benini}}, \bibinfo {author} {\bibfnamefont {D.}~\bibnamefont {Gang}}, \ and\ \bibinfo {author} {\bibfnamefont {L.~A.}\ \bibnamefont {Pando~Zayas}},\ }\href {\doibase 10.1007/JHEP03(2020)057} {\bibfield  {journal} {\bibinfo  {journal} {JHEP}\ }\textbf {\bibinfo {volume} {03}},\ \bibinfo {pages} {057} (\bibinfo {year} {2020})},\ \Eprint {http://arxiv.org/abs/1909.11612} {arXiv:1909.11612 [hep-th]} \BibitemShut {NoStop}%
\bibitem [{\citenamefont {David}\ \emph {et~al.}(2023)\citenamefont {David}, \citenamefont {Gava}, \citenamefont {Gupta},\ and\ \citenamefont {Narain}}]{David:2023btq}%
  \BibitemOpen
  \bibfield  {author} {\bibinfo {author} {\bibfnamefont {J.~R.}\ \bibnamefont {David}}, \bibinfo {author} {\bibfnamefont {E.}~\bibnamefont {Gava}}, \bibinfo {author} {\bibfnamefont {R.~K.}\ \bibnamefont {Gupta}}, \ and\ \bibinfo {author} {\bibfnamefont {K.~S.}\ \bibnamefont {Narain}},\ }\href {\doibase 10.1007/JHEP09(2023)171} {\bibfield  {journal} {\bibinfo  {journal} {JHEP}\ }\textbf {\bibinfo {volume} {09}},\ \bibinfo {pages} {171} (\bibinfo {year} {2023})},\ \Eprint {http://arxiv.org/abs/2303.14354} {arXiv:2303.14354 [hep-th]} \BibitemShut {NoStop}%
\bibitem [{\citenamefont {Gonz\'alez~Lezcano}\ \emph {et~al.}(2023)\citenamefont {Gonz\'alez~Lezcano}, \citenamefont {Jeon},\ and\ \citenamefont {Ray}}]{GonzalezLezcano:2023uar}%
  \BibitemOpen
  \bibfield  {author} {\bibinfo {author} {\bibfnamefont {A.}~\bibnamefont {Gonz\'alez~Lezcano}}, \bibinfo {author} {\bibfnamefont {I.}~\bibnamefont {Jeon}}, \ and\ \bibinfo {author} {\bibfnamefont {A.}~\bibnamefont {Ray}},\ }\href {\doibase 10.1103/PhysRevD.108.045018} {\bibfield  {journal} {\bibinfo  {journal} {Phys. Rev. D}\ }\textbf {\bibinfo {volume} {108}},\ \bibinfo {pages} {045018} (\bibinfo {year} {2023})},\ \Eprint {http://arxiv.org/abs/2305.12925} {arXiv:2305.12925 [hep-th]} \BibitemShut {NoStop}%
\bibitem [{\citenamefont {H.}\ \emph {et~al.}(2024)\citenamefont {H.}, \citenamefont {Athira}, \citenamefont {Chowdhury},\ and\ \citenamefont {Sen}}]{H:2023qko}%
  \BibitemOpen
  \bibfield  {author} {\bibinfo {author} {\bibfnamefont {A.~A.}\ \bibnamefont {H.}}, \bibinfo {author} {\bibfnamefont {P.~V.}\ \bibnamefont {Athira}}, \bibinfo {author} {\bibfnamefont {C.}~\bibnamefont {Chowdhury}}, \ and\ \bibinfo {author} {\bibfnamefont {A.}~\bibnamefont {Sen}},\ }\href {\doibase 10.1007/JHEP03(2024)095} {\bibfield  {journal} {\bibinfo  {journal} {JHEP}\ }\textbf {\bibinfo {volume} {03}},\ \bibinfo {pages} {095} (\bibinfo {year} {2024})},\ \Eprint {http://arxiv.org/abs/2306.07322} {arXiv:2306.07322 [hep-th]} \BibitemShut {NoStop}%
\bibitem [{\citenamefont {Iliesiu}\ and\ \citenamefont {Turiaci}(2021)}]{Iliesiu:2020qvm}%
  \BibitemOpen
  \bibfield  {author} {\bibinfo {author} {\bibfnamefont {L.~V.}\ \bibnamefont {Iliesiu}}\ and\ \bibinfo {author} {\bibfnamefont {G.~J.}\ \bibnamefont {Turiaci}},\ }\href {\doibase 10.1007/JHEP05(2021)145} {\bibfield  {journal} {\bibinfo  {journal} {JHEP}\ }\textbf {\bibinfo {volume} {05}},\ \bibinfo {pages} {145} (\bibinfo {year} {2021})},\ \Eprint {http://arxiv.org/abs/2003.02860} {arXiv:2003.02860 [hep-th]} \BibitemShut {NoStop}%
\bibitem [{\citenamefont {Kapec}\ \emph {et~al.}(2024{\natexlab{a}})\citenamefont {Kapec}, \citenamefont {Sheta}, \citenamefont {Strominger},\ and\ \citenamefont {Toldo}}]{Kapec:2023ruw}%
  \BibitemOpen
  \bibfield  {author} {\bibinfo {author} {\bibfnamefont {D.}~\bibnamefont {Kapec}}, \bibinfo {author} {\bibfnamefont {A.}~\bibnamefont {Sheta}}, \bibinfo {author} {\bibfnamefont {A.}~\bibnamefont {Strominger}}, \ and\ \bibinfo {author} {\bibfnamefont {C.}~\bibnamefont {Toldo}},\ }\href {\doibase 10.1103/PhysRevLett.133.021601} {\bibfield  {journal} {\bibinfo  {journal} {Phys. Rev. Lett.}\ }\textbf {\bibinfo {volume} {133}},\ \bibinfo {pages} {021601} (\bibinfo {year} {2024}{\natexlab{a}})},\ \Eprint {http://arxiv.org/abs/2310.00848} {arXiv:2310.00848 [hep-th]} \BibitemShut {NoStop}%
\bibitem [{\citenamefont {Rakic}\ \emph {et~al.}(2024)\citenamefont {Rakic}, \citenamefont {Rangamani},\ and\ \citenamefont {Turiaci}}]{Rakic:2023vhv}%
  \BibitemOpen
  \bibfield  {author} {\bibinfo {author} {\bibfnamefont {I.}~\bibnamefont {Rakic}}, \bibinfo {author} {\bibfnamefont {M.}~\bibnamefont {Rangamani}}, \ and\ \bibinfo {author} {\bibfnamefont {G.~J.}\ \bibnamefont {Turiaci}},\ }\href {\doibase 10.1007/JHEP06(2024)011} {\bibfield  {journal} {\bibinfo  {journal} {JHEP}\ }\textbf {\bibinfo {volume} {06}},\ \bibinfo {pages} {011} (\bibinfo {year} {2024})},\ \Eprint {http://arxiv.org/abs/2310.04532} {arXiv:2310.04532 [hep-th]} \BibitemShut {NoStop}%
\bibitem [{\citenamefont {Banerjee}\ \emph {et~al.}(2024)\citenamefont {Banerjee}, \citenamefont {Saha},\ and\ \citenamefont {Srinivasan}}]{Banerjee:2023gll}%
  \BibitemOpen
  \bibfield  {author} {\bibinfo {author} {\bibfnamefont {N.}~\bibnamefont {Banerjee}}, \bibinfo {author} {\bibfnamefont {M.}~\bibnamefont {Saha}}, \ and\ \bibinfo {author} {\bibfnamefont {S.}~\bibnamefont {Srinivasan}},\ }\href {\doibase 10.1007/JHEP02(2024)077} {\bibfield  {journal} {\bibinfo  {journal} {JHEP}\ }\textbf {\bibinfo {volume} {2024}},\ \bibinfo {pages} {077} (\bibinfo {year} {2024})},\ \Eprint {http://arxiv.org/abs/2311.09595} {arXiv:2311.09595 [hep-th]} \BibitemShut {NoStop}%
\bibitem [{\citenamefont {Maulik}\ \emph {et~al.}(2024)\citenamefont {Maulik}, \citenamefont {Pando~Zayas}, \citenamefont {Ray},\ and\ \citenamefont {Zhang}}]{Maulik:2024dwq}%
  \BibitemOpen
  \bibfield  {author} {\bibinfo {author} {\bibfnamefont {S.}~\bibnamefont {Maulik}}, \bibinfo {author} {\bibfnamefont {L.~A.}\ \bibnamefont {Pando~Zayas}}, \bibinfo {author} {\bibfnamefont {A.}~\bibnamefont {Ray}}, \ and\ \bibinfo {author} {\bibfnamefont {J.}~\bibnamefont {Zhang}},\ }\href {\doibase 10.1007/JHEP06(2024)034} {\bibfield  {journal} {\bibinfo  {journal} {JHEP}\ }\textbf {\bibinfo {volume} {06}},\ \bibinfo {pages} {034} (\bibinfo {year} {2024})},\ \Eprint {http://arxiv.org/abs/2401.16507} {arXiv:2401.16507 [hep-th]} \BibitemShut {NoStop}%
\bibitem [{\citenamefont {Kapec}\ \emph {et~al.}(2024{\natexlab{b}})\citenamefont {Kapec}, \citenamefont {Law},\ and\ \citenamefont {Toldo}}]{Kapec:2024zdj}%
  \BibitemOpen
  \bibfield  {author} {\bibinfo {author} {\bibfnamefont {D.}~\bibnamefont {Kapec}}, \bibinfo {author} {\bibfnamefont {Y.~T.~A.}\ \bibnamefont {Law}}, \ and\ \bibinfo {author} {\bibfnamefont {C.}~\bibnamefont {Toldo}},\ }\href@noop {} {\  (\bibinfo {year} {2024}{\natexlab{b}})},\ \Eprint {http://arxiv.org/abs/2409.14928} {arXiv:2409.14928 [hep-th]} \BibitemShut {NoStop}%
\bibitem [{\citenamefont {Kolanowski}\ \emph {et~al.}(2024)\citenamefont {Kolanowski}, \citenamefont {Marolf}, \citenamefont {Rakic}, \citenamefont {Rangamani},\ and\ \citenamefont {Turiaci}}]{Kolanowski:2024zrq}%
  \BibitemOpen
  \bibfield  {author} {\bibinfo {author} {\bibfnamefont {M.}~\bibnamefont {Kolanowski}}, \bibinfo {author} {\bibfnamefont {D.}~\bibnamefont {Marolf}}, \bibinfo {author} {\bibfnamefont {I.}~\bibnamefont {Rakic}}, \bibinfo {author} {\bibfnamefont {M.}~\bibnamefont {Rangamani}}, \ and\ \bibinfo {author} {\bibfnamefont {G.~J.}\ \bibnamefont {Turiaci}},\ }\href@noop {} {\  (\bibinfo {year} {2024})},\ \Eprint {http://arxiv.org/abs/2409.16248} {arXiv:2409.16248 [hep-th]} \BibitemShut {NoStop}%
\bibitem [{\citenamefont {Arnaudo}\ \emph {et~al.}(2024{\natexlab{a}})\citenamefont {Arnaudo}, \citenamefont {Bonelli},\ and\ \citenamefont {Tanzini}}]{PhysRevD.110.106006}%
  \BibitemOpen
  \bibfield  {author} {\bibinfo {author} {\bibfnamefont {P.}~\bibnamefont {Arnaudo}}, \bibinfo {author} {\bibfnamefont {G.}~\bibnamefont {Bonelli}}, \ and\ \bibinfo {author} {\bibfnamefont {A.}~\bibnamefont {Tanzini}},\ }\href {\doibase 10.1103/PhysRevD.110.106006} {\bibfield  {journal} {\bibinfo  {journal} {Phys. Rev. D}\ }\textbf {\bibinfo {volume} {110}},\ \bibinfo {pages} {106006} (\bibinfo {year} {2024}{\natexlab{a}})}\BibitemShut {NoStop}%
\bibitem [{\citenamefont {Dunne}(2008)}]{Dunne:2007rt}%
  \BibitemOpen
  \bibfield  {author} {\bibinfo {author} {\bibfnamefont {G.~V.}\ \bibnamefont {Dunne}},\ }\href {\doibase 10.1088/1751-8113/41/30/304006} {\bibfield  {journal} {\bibinfo  {journal} {J. Phys. A}\ }\textbf {\bibinfo {volume} {41}},\ \bibinfo {pages} {304006} (\bibinfo {year} {2008})},\ \Eprint {http://arxiv.org/abs/0711.1178} {arXiv:0711.1178 [hep-th]} \BibitemShut {NoStop}%
\bibitem [{\citenamefont {Denef}\ \emph {et~al.}(2010)\citenamefont {Denef}, \citenamefont {Hartnoll},\ and\ \citenamefont {Sachdev}}]{Denef:2009kn}%
  \BibitemOpen
  \bibfield  {author} {\bibinfo {author} {\bibfnamefont {F.}~\bibnamefont {Denef}}, \bibinfo {author} {\bibfnamefont {S.~A.}\ \bibnamefont {Hartnoll}}, \ and\ \bibinfo {author} {\bibfnamefont {S.}~\bibnamefont {Sachdev}},\ }\href {\doibase 10.1088/0264-9381/27/12/125001} {\bibfield  {journal} {\bibinfo  {journal} {Class. Quant. Grav.}\ }\textbf {\bibinfo {volume} {27}},\ \bibinfo {pages} {125001} (\bibinfo {year} {2010})},\ \Eprint {http://arxiv.org/abs/0908.2657} {arXiv:0908.2657 [hep-th]} \BibitemShut {NoStop}%
\bibitem [{\citenamefont {Aminov}\ \emph {et~al.}(2022)\citenamefont {Aminov}, \citenamefont {Grassi},\ and\ \citenamefont {Hatsuda}}]{Aminov:2020yma}%
  \BibitemOpen
  \bibfield  {author} {\bibinfo {author} {\bibfnamefont {G.}~\bibnamefont {Aminov}}, \bibinfo {author} {\bibfnamefont {A.}~\bibnamefont {Grassi}}, \ and\ \bibinfo {author} {\bibfnamefont {Y.}~\bibnamefont {Hatsuda}},\ }\href {\doibase 10.1007/s00023-021-01137-x} {\bibfield  {journal} {\bibinfo  {journal} {Annales Henri Poincare}\ }\textbf {\bibinfo {volume} {23}},\ \bibinfo {pages} {1951} (\bibinfo {year} {2022})},\ \Eprint {http://arxiv.org/abs/2006.06111} {arXiv:2006.06111 [hep-th]} \BibitemShut {NoStop}%
\bibitem [{\citenamefont {Bonelli}\ \emph {et~al.}(2022)\citenamefont {Bonelli}, \citenamefont {Iossa}, \citenamefont {Lichtig},\ and\ \citenamefont {Tanzini}}]{Bonelli:2021uvf}%
  \BibitemOpen
  \bibfield  {author} {\bibinfo {author} {\bibfnamefont {G.}~\bibnamefont {Bonelli}}, \bibinfo {author} {\bibfnamefont {C.}~\bibnamefont {Iossa}}, \bibinfo {author} {\bibfnamefont {D.~P.}\ \bibnamefont {Lichtig}}, \ and\ \bibinfo {author} {\bibfnamefont {A.}~\bibnamefont {Tanzini}},\ }\href {\doibase 10.1103/PhysRevD.105.044047} {\bibfield  {journal} {\bibinfo  {journal} {Phys. Rev. D}\ }\textbf {\bibinfo {volume} {105}},\ \bibinfo {pages} {044047} (\bibinfo {year} {2022})},\ \Eprint {http://arxiv.org/abs/2105.04483} {arXiv:2105.04483 [hep-th]} \BibitemShut {NoStop}%
\bibitem [{\citenamefont {Alday}\ \emph {et~al.}(2010)\citenamefont {Alday}, \citenamefont {Gaiotto},\ and\ \citenamefont {Tachikawa}}]{Alday:2009aq}%
  \BibitemOpen
  \bibfield  {author} {\bibinfo {author} {\bibfnamefont {L.~F.}\ \bibnamefont {Alday}}, \bibinfo {author} {\bibfnamefont {D.}~\bibnamefont {Gaiotto}}, \ and\ \bibinfo {author} {\bibfnamefont {Y.}~\bibnamefont {Tachikawa}},\ }\href {\doibase 10.1007/s11005-010-0369-5} {\bibfield  {journal} {\bibinfo  {journal} {Lett. Math. Phys.}\ }\textbf {\bibinfo {volume} {91}},\ \bibinfo {pages} {167} (\bibinfo {year} {2010})},\ \Eprint {http://arxiv.org/abs/0906.3219} {arXiv:0906.3219 [hep-th]} \BibitemShut {NoStop}%
\bibitem [{\citenamefont {Banerjee}\ and\ \citenamefont {Saha}(2023)}]{Banerjee:2023quv}%
  \BibitemOpen
  \bibfield  {author} {\bibinfo {author} {\bibfnamefont {N.}~\bibnamefont {Banerjee}}\ and\ \bibinfo {author} {\bibfnamefont {M.}~\bibnamefont {Saha}},\ }\href {\doibase 10.1007/JHEP07(2023)010} {\bibfield  {journal} {\bibinfo  {journal} {JHEP}\ }\textbf {\bibinfo {volume} {07}},\ \bibinfo {pages} {010} (\bibinfo {year} {2023})},\ \Eprint {http://arxiv.org/abs/2303.12415} {arXiv:2303.12415 [hep-th]} \BibitemShut {NoStop}%
\bibitem [{\citenamefont {Bautista}\ \emph {et~al.}(2024)\citenamefont {Bautista}, \citenamefont {Bonelli}, \citenamefont {Iossa}, \citenamefont {Tanzini},\ and\ \citenamefont {Zhou}}]{Bautista:2023sdf}%
  \BibitemOpen
  \bibfield  {author} {\bibinfo {author} {\bibfnamefont {Y.~F.}\ \bibnamefont {Bautista}}, \bibinfo {author} {\bibfnamefont {G.}~\bibnamefont {Bonelli}}, \bibinfo {author} {\bibfnamefont {C.}~\bibnamefont {Iossa}}, \bibinfo {author} {\bibfnamefont {A.}~\bibnamefont {Tanzini}}, \ and\ \bibinfo {author} {\bibfnamefont {Z.}~\bibnamefont {Zhou}},\ }\href {\doibase 10.1103/PhysRevD.109.084071} {\bibfield  {journal} {\bibinfo  {journal} {Phys. Rev. D}\ }\textbf {\bibinfo {volume} {109}},\ \bibinfo {pages} {084071} (\bibinfo {year} {2024})},\ \Eprint {http://arxiv.org/abs/2312.05965} {arXiv:2312.05965 [hep-th]} \BibitemShut {NoStop}%
\bibitem [{\citenamefont {Novaes}\ \emph {et~al.}(2019)\citenamefont {Novaes}, \citenamefont {Marinho}, \citenamefont {Lencs\'es},\ and\ \citenamefont {Casals}}]{Novaes:2018fry}%
  \BibitemOpen
  \bibfield  {author} {\bibinfo {author} {\bibfnamefont {F.}~\bibnamefont {Novaes}}, \bibinfo {author} {\bibfnamefont {C.}~\bibnamefont {Marinho}}, \bibinfo {author} {\bibfnamefont {M.}~\bibnamefont {Lencs\'es}}, \ and\ \bibinfo {author} {\bibfnamefont {M.}~\bibnamefont {Casals}},\ }\href {\doibase 10.1007/JHEP05(2019)033} {\bibfield  {journal} {\bibinfo  {journal} {JHEP}\ }\textbf {\bibinfo {volume} {05}},\ \bibinfo {pages} {033} (\bibinfo {year} {2019})},\ \Eprint {http://arxiv.org/abs/1811.11912} {arXiv:1811.11912 [gr-qc]} \BibitemShut {NoStop}%
\bibitem [{\citenamefont {Carneiro~da Cunha}\ and\ \citenamefont {Cavalcante}(2020)}]{CarneirodaCunha:2019tia}%
  \BibitemOpen
  \bibfield  {author} {\bibinfo {author} {\bibfnamefont {B.}~\bibnamefont {Carneiro~da Cunha}}\ and\ \bibinfo {author} {\bibfnamefont {J.~a.~P.}\ \bibnamefont {Cavalcante}},\ }\href {\doibase 10.1103/PhysRevD.102.105013} {\bibfield  {journal} {\bibinfo  {journal} {Phys. Rev. D}\ }\textbf {\bibinfo {volume} {102}},\ \bibinfo {pages} {105013} (\bibinfo {year} {2020})},\ \Eprint {http://arxiv.org/abs/1906.10638} {arXiv:1906.10638 [hep-th]} \BibitemShut {NoStop}%
\bibitem [{\citenamefont {Bianchi}\ \emph {et~al.}(2022)\citenamefont {Bianchi}, \citenamefont {Consoli}, \citenamefont {Grillo},\ and\ \citenamefont {Morales}}]{Bianchi:2021xpr}%
  \BibitemOpen
  \bibfield  {author} {\bibinfo {author} {\bibfnamefont {M.}~\bibnamefont {Bianchi}}, \bibinfo {author} {\bibfnamefont {D.}~\bibnamefont {Consoli}}, \bibinfo {author} {\bibfnamefont {A.}~\bibnamefont {Grillo}}, \ and\ \bibinfo {author} {\bibfnamefont {J.~F.}\ \bibnamefont {Morales}},\ }\href {\doibase 10.1016/j.physletb.2021.136837} {\bibfield  {journal} {\bibinfo  {journal} {Phys. Lett. B}\ }\textbf {\bibinfo {volume} {824}},\ \bibinfo {pages} {136837} (\bibinfo {year} {2022})},\ \Eprint {http://arxiv.org/abs/2105.04245} {arXiv:2105.04245 [hep-th]} \BibitemShut {NoStop}%
\bibitem [{\citenamefont {Fioravanti}\ and\ \citenamefont {Gregori}(2021)}]{Fioravanti:2021dce}%
  \BibitemOpen
  \bibfield  {author} {\bibinfo {author} {\bibfnamefont {D.}~\bibnamefont {Fioravanti}}\ and\ \bibinfo {author} {\bibfnamefont {D.}~\bibnamefont {Gregori}},\ }\href@noop {} {\  (\bibinfo {year} {2021})},\ \Eprint {http://arxiv.org/abs/2112.11434} {arXiv:2112.11434 [hep-th]} \BibitemShut {NoStop}%
\bibitem [{\citenamefont {Aminov}\ \emph {et~al.}(2023)\citenamefont {Aminov}, \citenamefont {Arnaudo}, \citenamefont {Bonelli}, \citenamefont {Grassi},\ and\ \citenamefont {Tanzini}}]{Aminov:2023jve}%
  \BibitemOpen
  \bibfield  {author} {\bibinfo {author} {\bibfnamefont {G.}~\bibnamefont {Aminov}}, \bibinfo {author} {\bibfnamefont {P.}~\bibnamefont {Arnaudo}}, \bibinfo {author} {\bibfnamefont {G.}~\bibnamefont {Bonelli}}, \bibinfo {author} {\bibfnamefont {A.}~\bibnamefont {Grassi}}, \ and\ \bibinfo {author} {\bibfnamefont {A.}~\bibnamefont {Tanzini}},\ }\href {\doibase 10.1007/JHEP11(2023)059} {\bibfield  {journal} {\bibinfo  {journal} {JHEP}\ }\textbf {\bibinfo {volume} {11}},\ \bibinfo {pages} {059} (\bibinfo {year} {2023})},\ \Eprint {http://arxiv.org/abs/2307.10141} {arXiv:2307.10141 [hep-th]} \BibitemShut {NoStop}%
\bibitem [{\citenamefont {Aminov}\ and\ \citenamefont {Arnaudo}(2024)}]{Aminov:2024mul}%
  \BibitemOpen
  \bibfield  {author} {\bibinfo {author} {\bibfnamefont {G.}~\bibnamefont {Aminov}}\ and\ \bibinfo {author} {\bibfnamefont {P.}~\bibnamefont {Arnaudo}},\ }\href@noop {} {\  (\bibinfo {year} {2024})},\ \Eprint {http://arxiv.org/abs/2409.06681} {arXiv:2409.06681 [hep-th]} \BibitemShut {NoStop}%
\bibitem [{\citenamefont {Iliesiu}\ \emph {et~al.}(2022)\citenamefont {Iliesiu}, \citenamefont {Murthy},\ and\ \citenamefont {Turiaci}}]{Iliesiu:2022onk}%
  \BibitemOpen
  \bibfield  {author} {\bibinfo {author} {\bibfnamefont {L.~V.}\ \bibnamefont {Iliesiu}}, \bibinfo {author} {\bibfnamefont {S.}~\bibnamefont {Murthy}}, \ and\ \bibinfo {author} {\bibfnamefont {G.~J.}\ \bibnamefont {Turiaci}},\ }\href@noop {} {\  (\bibinfo {year} {2022})},\ \Eprint {http://arxiv.org/abs/2209.13608} {arXiv:2209.13608 [hep-th]} \BibitemShut {NoStop}%
\bibitem [{\citenamefont {Teukolsky}(1972)}]{PhysRevLett.29.1114}%
  \BibitemOpen
  \bibfield  {author} {\bibinfo {author} {\bibfnamefont {S.~A.}\ \bibnamefont {Teukolsky}},\ }\href {\doibase 10.1103/PhysRevLett.29.1114} {\bibfield  {journal} {\bibinfo  {journal} {Phys. Rev. Lett.}\ }\textbf {\bibinfo {volume} {29}},\ \bibinfo {pages} {1114} (\bibinfo {year} {1972})}\BibitemShut {NoStop}%
\bibitem [{Note1()}]{Note1}%
  \BibitemOpen
  \bibinfo {note} {This is a convergent series \cite {Arnaudo:2022ivo} whose coefficients can be computed explicitly, see \cite {Aminov:2020yma,Bonelli:2021uvf} for details.}\BibitemShut {Stop}%
\bibitem [{Note2()}]{Note2}%
  \BibitemOpen
  \bibinfo {note} {These quantities $\mu _1$ and $\mu _2$, together with the parameter $\mu $, are the masses of the hypermultiplets in the fundamental representation from the gauge theory point of view.}\BibitemShut {Stop}%
\bibitem [{\citenamefont {Bonelli}\ \emph {et~al.}(2023)\citenamefont {Bonelli}, \citenamefont {Iossa}, \citenamefont {Panea~Lichtig},\ and\ \citenamefont {Tanzini}}]{Bonelli:2022ten}%
  \BibitemOpen
  \bibfield  {author} {\bibinfo {author} {\bibfnamefont {G.}~\bibnamefont {Bonelli}}, \bibinfo {author} {\bibfnamefont {C.}~\bibnamefont {Iossa}}, \bibinfo {author} {\bibfnamefont {D.}~\bibnamefont {Panea~Lichtig}}, \ and\ \bibinfo {author} {\bibfnamefont {A.}~\bibnamefont {Tanzini}},\ }\href {\doibase 10.1007/s00220-022-04497-5} {\bibfield  {journal} {\bibinfo  {journal} {Commun. Math. Phys.}\ }\textbf {\bibinfo {volume} {397}},\ \bibinfo {pages} {635} (\bibinfo {year} {2023})},\ \Eprint {http://arxiv.org/abs/2201.04491} {arXiv:2201.04491 [hep-th]} \BibitemShut {NoStop}%
\bibitem [{Note3()}]{Note3}%
  \BibitemOpen
  \bibinfo {note} {Strictly speaking the GY formula computes the regularised determinant without zero modes ${\protect \rm det}'$. For $s=1,2$ the operator we consider do not display zero modes so that ${\protect \rm det}'={\protect \rm det}$.}\BibitemShut {Stop}%
\bibitem [{Note4()}]{Note4}%
  \BibitemOpen
  \bibinfo {note} {From the gauge theory viewpoint, this procedure corresponds to the holomorphic decoupling of the hypermultiplet mass $\mu _2$, and it produces the $N_f=2$ theory from the original $N_f=3$ one. The new parameter $\Lambda \equiv -8 M^2 \omega (\omega -m\protect \,\Omega _H^{\protect \text {ext}})$, with $\Omega _H^{\protect \text {ext}}=1/(2M)$, corresponds to the instanton counting parameter of the $N_f=2$ theory.}\BibitemShut {Stop}%
\bibitem [{\citenamefont {Cavalcante}\ \emph {et~al.}(2024)\citenamefont {Cavalcante}, \citenamefont {Richartz},\ and\ \citenamefont {da~Cunha}}]{Cavalcante:2024kmy}%
  \BibitemOpen
  \bibfield  {author} {\bibinfo {author} {\bibfnamefont {J.~a.~P.}\ \bibnamefont {Cavalcante}}, \bibinfo {author} {\bibfnamefont {M.}~\bibnamefont {Richartz}}, \ and\ \bibinfo {author} {\bibfnamefont {B.~C.}\ \bibnamefont {da~Cunha}},\ }\href@noop {} {\  (\bibinfo {year} {2024})},\ \Eprint {http://arxiv.org/abs/2408.13964} {arXiv:2408.13964 [gr-qc]} \BibitemShut {NoStop}%
\bibitem [{\citenamefont {Yang}\ \emph {et~al.}(2013)\citenamefont {Yang}, \citenamefont {Zimmerman}, \citenamefont {Zengino\u{g}lu}, \citenamefont {Zhang}, \citenamefont {Berti},\ and\ \citenamefont {Chen}}]{Yang:2013uba}%
  \BibitemOpen
  \bibfield  {author} {\bibinfo {author} {\bibfnamefont {H.}~\bibnamefont {Yang}}, \bibinfo {author} {\bibfnamefont {A.}~\bibnamefont {Zimmerman}}, \bibinfo {author} {\bibfnamefont {A.}~\bibnamefont {Zengino\u{g}lu}}, \bibinfo {author} {\bibfnamefont {F.}~\bibnamefont {Zhang}}, \bibinfo {author} {\bibfnamefont {E.}~\bibnamefont {Berti}}, \ and\ \bibinfo {author} {\bibfnamefont {Y.}~\bibnamefont {Chen}},\ }\href {\doibase 10.1103/PhysRevD.88.044047} {\bibfield  {journal} {\bibinfo  {journal} {Phys. Rev. D}\ }\textbf {\bibinfo {volume} {88}},\ \bibinfo {pages} {044047} (\bibinfo {year} {2013})},\ \Eprint {http://arxiv.org/abs/1307.8086} {arXiv:1307.8086 [gr-qc]} \BibitemShut {NoStop}%
\bibitem [{\citenamefont {Casals}\ and\ \citenamefont {Longo~Micchi}(2019)}]{Casals:2019vdb}%
  \BibitemOpen
  \bibfield  {author} {\bibinfo {author} {\bibfnamefont {M.}~\bibnamefont {Casals}}\ and\ \bibinfo {author} {\bibfnamefont {L.~F.}\ \bibnamefont {Longo~Micchi}},\ }\href {\doibase 10.1103/PhysRevD.99.084047} {\bibfield  {journal} {\bibinfo  {journal} {Phys. Rev. D}\ }\textbf {\bibinfo {volume} {99}},\ \bibinfo {pages} {084047} (\bibinfo {year} {2019})},\ \Eprint {http://arxiv.org/abs/1901.04586} {arXiv:1901.04586 [gr-qc]} \BibitemShut {NoStop}%
\bibitem [{\citenamefont {Berti}\ \emph {et~al.}(2006)\citenamefont {Berti}, \citenamefont {Cardoso},\ and\ \citenamefont {Casals}}]{Berti:2005gp}%
  \BibitemOpen
  \bibfield  {author} {\bibinfo {author} {\bibfnamefont {E.}~\bibnamefont {Berti}}, \bibinfo {author} {\bibfnamefont {V.}~\bibnamefont {Cardoso}}, \ and\ \bibinfo {author} {\bibfnamefont {M.}~\bibnamefont {Casals}},\ }\href {\doibase 10.1103/PhysRevD.73.109902} {\bibfield  {journal} {\bibinfo  {journal} {Phys. Rev. D}\ }\textbf {\bibinfo {volume} {73}},\ \bibinfo {pages} {024013} (\bibinfo {year} {2006})},\ \bibinfo {note} {[Erratum: Phys.Rev.D 73, 109902 (2006)]},\ \Eprint {http://arxiv.org/abs/gr-qc/0511111} {arXiv:gr-qc/0511111} \BibitemShut {NoStop}%
\bibitem [{\citenamefont {Hod}(2013)}]{Hod:2013fea}%
  \BibitemOpen
  \bibfield  {author} {\bibinfo {author} {\bibfnamefont {S.}~\bibnamefont {Hod}},\ }\href {\doibase 10.1103/PhysRevD.88.084018} {\bibfield  {journal} {\bibinfo  {journal} {Phys. Rev. D}\ }\textbf {\bibinfo {volume} {88}},\ \bibinfo {pages} {084018} (\bibinfo {year} {2013})},\ \Eprint {http://arxiv.org/abs/1311.3007} {arXiv:1311.3007 [gr-qc]} \BibitemShut {NoStop}%
\bibitem [{Note5()}]{Note5}%
  \BibitemOpen
  \bibinfo {note} {In detail, this extra factor can be seen as the combination of an overall factor $\epsilon ^{\protect \frac {1}{2}-\mu }$ in the normalization of the irregular semiclassical block around infinity and a factor $\epsilon ^a$ coming from the irregular semiclassical block around zero with shifted intermediate momentum, arising in the confluence procedure between the singularities at $z=0$ and $z=1$. Specifically, the first factor can be seen by taking the semiclassical limit of the $\Lambda $-factors in the first line of formula (3.2.7) in \cite {Bonelli:2022ten}. In particular, considering the semiclassical limit $b\to 0$, the term $b\protect \,\Lambda $ corresponds to our parameter $\epsilon $, and the semiclassical limit of $\Delta _{2,1}-\theta \protect \,b\protect \,\mu $ in \cite {Bonelli:2022ten} becomes $-\protect \frac {1}{2}+\mu $ in our notation. This, consistently with our choices of the local solution around $z=\infty $, cancels the factor $\epsilon ^{\protect \frac {1}{2}-\mu }$.
  The other factor can be seen by taking the semiclassical limit of the $\Lambda _2$-term in formula (3.4.4) in \cite {Bonelli:2022ten}, when considering the shift of the intermediate momentum. Indeed, $\Delta _{\sigma }\to \sigma \protect \,a$ in the semiclassical limit, and this simplifies with the corresponding factor in the connection formula \protect \eqref {connectioninftyto1}. This, in turn, implies the simplification of the term $\epsilon ^a$ in \protect \eqref {detfactorized} after taking the confluent limit.}\BibitemShut {Stop}%
\bibitem [{\citenamefont {Castro}\ \emph {et~al.}(2017)\citenamefont {Castro}, \citenamefont {Keeler},\ and\ \citenamefont {Szepietowski}}]{Castro:2017mfj}%
  \BibitemOpen
  \bibfield  {author} {\bibinfo {author} {\bibfnamefont {A.}~\bibnamefont {Castro}}, \bibinfo {author} {\bibfnamefont {C.}~\bibnamefont {Keeler}}, \ and\ \bibinfo {author} {\bibfnamefont {P.}~\bibnamefont {Szepietowski}},\ }\href {\doibase 10.1007/JHEP10(2017)070} {\bibfield  {journal} {\bibinfo  {journal} {JHEP}\ }\textbf {\bibinfo {volume} {10}},\ \bibinfo {pages} {070} (\bibinfo {year} {2017})},\ \Eprint {http://arxiv.org/abs/1707.06245} {arXiv:1707.06245 [hep-th]} \BibitemShut {NoStop}%
\bibitem [{Note6()}]{Note6}%
  \BibitemOpen
  \bibinfo {note} {The Matsubara frequencies associated with the anti-QNMs are obtained by taking the asymptotic behaviour $\psi _+^{(1)}$ in \protect \eqref {localsolutionsat1} which, together with the corresponding normalization, amounts to impose $2a_1=k$, $k\ge 0$, so that $\omega _k^{(M)}(-s)=\protect \overline {\omega }_k^{(M)}(s)$.}\BibitemShut {Stop}%
\bibitem [{\citenamefont {Arnaudo}\ \emph {et~al.}(2024{\natexlab{b}})\citenamefont {Arnaudo}, \citenamefont {Bonelli},\ and\ \citenamefont {Tanzini}}]{Arnaudo:2022ivo}%
  \BibitemOpen
  \bibfield  {author} {\bibinfo {author} {\bibfnamefont {P.}~\bibnamefont {Arnaudo}}, \bibinfo {author} {\bibfnamefont {G.}~\bibnamefont {Bonelli}}, \ and\ \bibinfo {author} {\bibfnamefont {A.}~\bibnamefont {Tanzini}},\ }\href {\doibase 10.1007/s00023-023-01349-3} {\bibfield  {journal} {\bibinfo  {journal} {Annales Henri Poincare}\ }\textbf {\bibinfo {volume} {25}},\ \bibinfo {pages} {2389} (\bibinfo {year} {2024}{\natexlab{b}})},\ \Eprint {http://arxiv.org/abs/2212.06741} {arXiv:2212.06741 [hep-th]} \BibitemShut {NoStop}%
\bibitem [{Note7()}]{Note7}%
  \BibitemOpen
  \bibinfo {note} {The relevant connection formula for the Whittaker functions from the selected solution at $z=\infty $ reads \begin {equation*} W_{\mu ,-a_1}(\epsilon \protect \,z -\epsilon )\sim \protect \frac {\Gamma \left (-2a_1\right )}{\Gamma \left (\protect \frac {1}{2}-a_1-\mu \right )}\psi _+^{(1)}(z)+\protect \frac {\Gamma \left (2a_1\right )}{\Gamma \left (\protect \frac {1}{2}+a_1-\mu \right )}\psi _-^{(1)}(z). \end {equation*}}\BibitemShut {NoStop}%
\bibitem [{\citenamefont {Wang}\ and\ \citenamefont {Guo}(1989)}]{wangguo}%
  \BibitemOpen
  \bibfield  {author} {\bibinfo {author} {\bibfnamefont {Z.~X.}\ \bibnamefont {Wang}}\ and\ \bibinfo {author} {\bibfnamefont {D.~R.}\ \bibnamefont {Guo}},\ }\href {\doibase 10.1142/0653} {\emph {\bibinfo {title} {Special Functions}}}\ (\bibinfo  {publisher} {WORLD SCIENTIFIC},\ \bibinfo {year} {1989})\ \Eprint {http://arxiv.org/abs/https://www.worldscientific.com/doi/pdf/10.1142/0653} {https://www.worldscientific.com/doi/pdf/10.1142/0653} \BibitemShut {NoStop}%
\bibitem [{\citenamefont {Kurokawa}\ \emph {et~al.}(2012)\citenamefont {Kurokawa}, \citenamefont {Kurihara},\ and\ \citenamefont {Saito}}]{iwasawa}%
  \BibitemOpen
  \bibfield  {author} {\bibinfo {author} {\bibfnamefont {N.}~\bibnamefont {Kurokawa}}, \bibinfo {author} {\bibfnamefont {M.}~\bibnamefont {Kurihara}}, \ and\ \bibinfo {author} {\bibfnamefont {T.}~\bibnamefont {Saito}},\ }\href@noop {} {\emph {\bibinfo {title} {Number theory, 3, Iwasawa theory and modular forms}}}\ (\bibinfo  {publisher} {Translations of Mathematical Monographs, 242, Iwanami Series in Modern Mathematics, American Mathematical Society, Providence, RI, 20},\ \bibinfo {year} {2012})\BibitemShut {NoStop}%
\bibitem [{\citenamefont {Gaiotto}(2013)}]{Gaiotto:2009ma}%
  \BibitemOpen
  \bibfield  {author} {\bibinfo {author} {\bibfnamefont {D.}~\bibnamefont {Gaiotto}},\ }\href {\doibase 10.1088/1742-6596/462/1/012014} {\bibfield  {journal} {\bibinfo  {journal} {J. Phys. Conf. Ser.}\ }\textbf {\bibinfo {volume} {462}},\ \bibinfo {pages} {012014} (\bibinfo {year} {2013})},\ \Eprint {http://arxiv.org/abs/0908.0307} {arXiv:0908.0307 [hep-th]} \BibitemShut {NoStop}%
\bibitem [{\citenamefont {Bonelli}\ \emph {et~al.}(2012)\citenamefont {Bonelli}, \citenamefont {Maruyoshi},\ and\ \citenamefont {Tanzini}}]{Bonelli:2011aa}%
  \BibitemOpen
  \bibfield  {author} {\bibinfo {author} {\bibfnamefont {G.}~\bibnamefont {Bonelli}}, \bibinfo {author} {\bibfnamefont {K.}~\bibnamefont {Maruyoshi}}, \ and\ \bibinfo {author} {\bibfnamefont {A.}~\bibnamefont {Tanzini}},\ }\href {\doibase 10.1007/JHEP02(2012)031} {\bibfield  {journal} {\bibinfo  {journal} {JHEP}\ }\textbf {\bibinfo {volume} {02}},\ \bibinfo {pages} {031} (\bibinfo {year} {2012})},\ \Eprint {http://arxiv.org/abs/1112.1691} {arXiv:1112.1691 [hep-th]} \BibitemShut {NoStop}%
\bibitem [{\citenamefont {Gaiotto}\ and\ \citenamefont {Teschner}(2012)}]{Gaiotto:2012sf}%
  \BibitemOpen
  \bibfield  {author} {\bibinfo {author} {\bibfnamefont {D.}~\bibnamefont {Gaiotto}}\ and\ \bibinfo {author} {\bibfnamefont {J.}~\bibnamefont {Teschner}},\ }\href {\doibase 10.1007/JHEP12(2012)050} {\bibfield  {journal} {\bibinfo  {journal} {JHEP}\ }\textbf {\bibinfo {volume} {12}},\ \bibinfo {pages} {050} (\bibinfo {year} {2012})},\ \Eprint {http://arxiv.org/abs/1203.1052} {arXiv:1203.1052 [hep-th]} \BibitemShut {NoStop}%
\bibitem [{\citenamefont {Belavin}\ \emph {et~al.}(1984)\citenamefont {Belavin}, \citenamefont {Polyakov},\ and\ \citenamefont {Zamolodchikov}}]{Belavin:1984vu}%
  \BibitemOpen
  \bibfield  {author} {\bibinfo {author} {\bibfnamefont {A.~A.}\ \bibnamefont {Belavin}}, \bibinfo {author} {\bibfnamefont {A.~M.}\ \bibnamefont {Polyakov}}, \ and\ \bibinfo {author} {\bibfnamefont {A.~B.}\ \bibnamefont {Zamolodchikov}},\ }\href {\doibase 10.1016/0550-3213(84)90052-X} {\bibfield  {journal} {\bibinfo  {journal} {Nucl. Phys. B}\ }\textbf {\bibinfo {volume} {241}},\ \bibinfo {pages} {333} (\bibinfo {year} {1984})}\BibitemShut {NoStop}%
\bibitem [{\citenamefont {Matone}(1995)}]{Matone:1995rx}%
  \BibitemOpen
  \bibfield  {author} {\bibinfo {author} {\bibfnamefont {M.}~\bibnamefont {Matone}},\ }\href {\doibase 10.1016/0370-2693(95)00920-G} {\bibfield  {journal} {\bibinfo  {journal} {Phys. Lett. B}\ }\textbf {\bibinfo {volume} {357}},\ \bibinfo {pages} {342} (\bibinfo {year} {1995})},\ \Eprint {http://arxiv.org/abs/hep-th/9506102} {arXiv:hep-th/9506102} \BibitemShut {NoStop}%
\bibitem [{\citenamefont {Datta}\ and\ \citenamefont {David}(2012)}]{Datta:2011za}%
  \BibitemOpen
  \bibfield  {author} {\bibinfo {author} {\bibfnamefont {S.}~\bibnamefont {Datta}}\ and\ \bibinfo {author} {\bibfnamefont {J.~R.}\ \bibnamefont {David}},\ }\href {\doibase 10.1007/JHEP03(2012)079} {\bibfield  {journal} {\bibinfo  {journal} {JHEP}\ }\textbf {\bibinfo {volume} {03}},\ \bibinfo {pages} {079} (\bibinfo {year} {2012})},\ \Eprint {http://arxiv.org/abs/1112.4619} {arXiv:1112.4619 [hep-th]} \BibitemShut {NoStop}%
\end{thebibliography}%

  \appendix
\section{ APPENDICES}
\subsection{Determinant of Confluent Heun differential operators}\label{appendixCHeunoverWhitt}

In this appendix we compute the determinant of the confluent Heun's operator $\mathcal{D}$ in \eqref{confluentHeun} by applying the modified version of the Gelfand-Yaglom (GY) theorem \cite{PhysRevD.110.106006}, and adapted to the presence of an irregular singularity, where the boundary condition is imposed. We recall that the GY theorem allows to compute the ratio of determinants of linear second-order differential operators with Dirichlet boundary conditions in terms of solutions of an auxiliary Cauchy problem with vanishing initial condition
and normalized first derivative.

A basis of local solutions of Eq.\eqref{confluentHeun} around $z=1$ reads
\begin{equation}\label{localsolutionsat1}
\psi_{\pm}^{(1)}(z)\sim (z-1)^{\frac{1}{2}\pm a_1}\left[1+\mathcal{O}\brc{z-1}\right].
\end{equation}
and around $z=\infty$
\begin{equation}\label{localsolutionsatinfty}
\psi_{\pm}^{(\infty)}(z)\sim e^{\pm\epsilon z/2}z^{\mp \mu}\left[1+\mathcal{O}\brc{1/z}\right].
\end{equation}
As for the boundary conditions, these depend on the signs of $\mathrm{Re}\brc{a_1}$ and $\mathrm{Re}\brc{\epsilon}$.
We remark that in our black hole problem, both $a_1$ and $\epsilon$ depend on $\omega$, and, in particular, the selected local solutions depend on the sign of $\mathrm{Im}(\omega)$.
In order to apply the Gelfand-Yaglom theorem in the regular singular case, the standard vanishing initial condition at $z=1$ is reformulated by asking the function $\psi$ to satisfy 
\begin{equation}\label{local1}
\lim_{z\to 1}\left((z-1)^{\frac{1}{2}+a_1}\right)^{-1}\psi(z)=0,
\end{equation}
where we supposed $\mathrm{Re}(a_1)<0$, selecting the local solution $\psi_{-}^{(1)}$.
The boundary condition at $z=\infty$ is instead reformulated by asking the function $\psi$ to satisfy 
\begin{equation}\label{local2}
\lim_{z\to \infty}e^{-\epsilon z/2}\psi(z)=0,
\end{equation}
where we supposed $\mathrm{Re}(\epsilon)>0$, selecting the local solution $\psi_{-}^{(\infty)}$.
We then have to define the reference problem. We require this to have (at least) one regular singular point and one irregular singularity of rank 1, with the same asymptotic behavior \eqref{localsolutionsat1} around $z=1$ and \eqref{localsolutionsatinfty} around $z=\infty$.
We modify the potential in \eqref{confluentHeun} by setting $a_0^2=\frac{1}{4}$ and $u=\frac{1}{4}-a_1^2+\mu\,\epsilon.$ This gives
\begin{equation}\label{Whittakerdiffop}
\tilde{\mathcal{D}}:\quad\frac{\mathrm{d}^2}{\mathrm{d}z^2} -\frac{4 a_1^2-4 \mu (z-1) \epsilon +z^2 \epsilon ^2-2 z \epsilon ^2+\epsilon ^2-1}{4 (z-1)^2},
\end{equation}
which is a Whittaker differential operator.
The basis of local solutions around $z=1$ and $z=\infty$ of this differential equation read precisely as in \eqref{localsolutionsat1} and \eqref{localsolutionsatinfty}, respectively.
To proceed with the application of the modified GY theorem, the strategy is to consider the associated problem (see Appendix A.1 in \cite{PhysRevD.110.106006}) as in the standard GY theorem, but take the normalized solution at a point close to $z=\infty$ in terms of the corresponding local solutions \eqref{localsolutionsatinfty}.
Then, using the connection matrix, it is possible to analytically continue the solution close to the point $z=1$ and evaluate it there. Considering the ratio with respect to the determinant of the reference problem and removing the cut-off, we get the ratio of connection coefficients as a final result.
For notational convenience, we denote with an index 1 the selected local solutions in \eqref{local1} and \eqref{local2} and with an index 2 the discarded ones.
The normalized solution around $z=\infty$ satisfying $u_{[\delta]}(1/\delta)=0$ and $u_{[\delta]}'(1/\delta)=1$
is given by 
\begin{equation}
u_{[\delta]}(z)=\frac{\psi_i^{(\infty)}(1/\delta)}{W(\psi_1^{(\infty)},\psi_2^{(\infty)})(1/\delta)}{\varepsilon_{ij}} \psi_j^{(\infty)}(z),
\end{equation}
where $\varepsilon_{ij}$ is the $2\times2$ anti-symmetric matrix with $\varepsilon_{12}=1$ and $W(f,g)=f\,g'-g\,f'$ is the Wronskian. 
Let us now analytically continue this solution to the neighborhood of $z=1$, as
\begin{equation}
\psi_i^{(\infty)}(z)=\mathcal{C}_{ij}\psi_j^{(1)}(z),
\end{equation}
and evaluate
\begin{equation}
u_{[\delta]}(1+\delta')=\frac{\psi_i^{(\infty)}(1/\delta)}{W(\psi_1^{(\infty)},\psi_2^{(\infty)})(1/\delta)}{\varepsilon_{ij}} \mathcal{C}_{jk}\psi_k^{(1)}(1+\delta')\, .
\end{equation}
By using the expansion of the local solutions around $z=1$ denoting with
\begin{equation}
\begin{aligned}
&\rho_1^{(1)}=\frac{1}{2}-a_1,\quad \rho_2^{(1)}=\frac{1}{2}+a_1,
\end{aligned}
\end{equation}
one finds
\begin{equation}
u_{[\delta]}(1+\delta')=
\frac{e^{(-1)^i\frac{\epsilon}{2\delta}}\delta^{(-1)^i\mu}
{\varepsilon_{ij}} \mathcal{C}_{jk}\brc{\delta'}^{\rho_k^{(1)}}}{W(\psi_1^{(\infty)},\psi_2^{(\infty)})(1/\delta)}
\left[1+\mathcal{O}\left(\delta,\delta'\right)\right].
\end{equation}
Consider now the ratio
\begin{equation}
\frac{u_{[\delta]}(1+\delta')}{\tilde{u}_{[\delta]}(1+\delta')},
\end{equation}
where the denominator is given by the above procedure applied to the reference operator $\tilde{\mathcal{D}}$ and with the same cut-off assignment.
As we remove the cut-off, in the limit $\delta,\delta'\to 0$ and using the assumptions $\mathrm{Re}(a_1)<0$ and $\mathrm{Re}(\epsilon)>0$, the leading order term is given by
\begin{equation}\label{GYi}
\begin{aligned}
\frac{\mathrm{det}({\mathcal{D}})}{\mathrm{det}({\tilde{\mathcal{D}}})}&=\lim_{\delta,\delta'\to0^+}
\frac{u_{[\delta]}(1+\delta')}{\tilde{u}_{[\delta]}(1+\delta')}\\
&=\lim_{\delta,\delta'\to 0^+}
\frac{-e^{\frac{\epsilon}{2\delta}}\delta^{\mu}\mathcal{C}_{12}\brc{\delta'}^{\frac{1}{2}+a_1}}{-e^{\frac{\epsilon}{2\delta}}\delta^{\mu}\tilde{\mathcal{C}}_{12}\brc{\delta'}^{\frac{1}{2}+a_1}}= \frac{\mathcal{C}_{12}}{\tilde{\mathcal{C}}_{12}}.
\end{aligned}
\end{equation}
By using the explicit formulae for the connection coefficients in \cite{Bonelli:2022ten}, the RHS of \eqref{GYi} reads \footnote{The relevant connection formula for the Whittaker functions from the selected solution at $z=\infty$ reads 
\begin{equation*}
W_{\mu,-a_1}(\epsilon\,z -\epsilon )\sim\frac{\Gamma\brc{-2a_1}}{\Gamma\brc{\frac{1}{2}-a_1-\mu}}\psi_+^{(1)}(z)+\frac{\Gamma\brc{2a_1}}{\Gamma\brc{\frac{1}{2}+a_1-\mu}}\psi_-^{(1)}(z).
\end{equation*}}
\begin{equation}\label{ratiodetpositive}
\begin{aligned}
&\frac{ \sum_{\sigma=\pm}\frac{\Gamma\brc{-2\sigma a}\Gamma\brc{1-2\sigma\,a}\Gamma\brc{-2a_1}\epsilon^{\frac{1}{2}-\mu+\sigma a}e^{-\frac{1}{2}\left(\sigma\partial_a+\partial_{a_1}+\partial_{\mu}\right)F}}{\Gamma\brc{\frac{1}{2}+\mu_1-\sigma a}\Gamma\brc{\frac{1}{2}+\mu_2-\sigma a}\Gamma\brc{\frac{1}{2}-\mu-\sigma a}}}{\frac{\Gamma\brc{-2a_1}}{\Gamma\brc{\frac{1}{2}-a_1-\mu}}}.
\end{aligned}
\end{equation}
Finally, to get the expression for $\det(\mathcal{D})$, we compute directly the determinant of the Whittaker operator \eqref{Whittakerdiffop} via the same argument as in Appendix C.2 in \cite{PhysRevD.110.106006}. First, we redefine the local variable as
\begin{equation}
y=\epsilon\,z-\epsilon,
\end{equation}
so that the differential operator becomes of the form of equation (4) page 300 in \cite{wangguo} with
\begin{equation}
k=\mu,\quad m=-a_1. 
\end{equation}
Then, redefining the operator by conjugating it with respect to 
\begin{equation}
e^{-y/2} y^{\frac{1}{2}-a_1},
\end{equation}
the differential operator $\tilde{\mathcal{D}}$ gets mapped into the operator
\begin{equation}
\tilde{\mathcal{D}}_1\colon\quad y\frac{\mathrm{d}^2}{\mathrm{d}\,y^2}+ (1-2 a_1-y)\frac{\mathrm{d}}{\mathrm{d}\,y}-\frac{1}{2}+a_1+\mu,
\end{equation}
which is a Kummer's operator in the form of equation (1) page 302 in \cite{wangguo} with 
\begin{equation}
\gamma=1-2a_1,\quad\quad  \alpha=\frac{1}{2}-a_1-\mu.
\end{equation}
As argued in \cite{PhysRevD.110.106006} these operations do not affect the determinant of the differential operator, and the boundary conditions are imposed at $y=0$ and $y=\infty$.
If we now consider the eigenvalue problem
\begin{equation}
\tilde{\mathcal{D}}_1\,\psi(y)=\lambda\,\psi(y),
\end{equation}
we have that the regular solution around $y=0$ is the confluent hypergeometric function \cite{wangguo}
\begin{equation}
F\left(\frac{1}{2}-a_1-\mu+\lambda,1-2 a_1,y\right).
\end{equation}
According to the connection formula (7) on page 314 of \cite{wangguo}, the discrete set of values for $\lambda$ which make the solution regular at $y=\infty$ satisfies
\begin{equation}
\frac{1}{2}-a_1-\mu+\lambda_n=-n,\quad n\in\mathbb{Z}_{\ge 0},
\end{equation}
that is 
\begin{equation}
\lambda_n=-n-\frac{1}{2}+a_1+\mu,\quad n\in\mathbb{Z}_{\ge 0}.
\end{equation}
Hence, denoting with a tilde the $\zeta$-regularization of the infinite product, the determinant of $\tilde{\mathcal{D}}_1$ (which is equal to the one of $\tilde{\mathcal{D}}$) is given by
\begin{equation}\label{detconfluentheun}
\mathrm{det}\tilde{\mathcal{D}}=\widetilde{\prod}_{n\ge 0}\left(-n-\frac{1}{2}+a_1+\mu\right)=\frac{\sqrt{2\,\pi}}{\Gamma\left(\frac{1}{2}-a_1-\mu\right)},
\end{equation}
where we used the $\zeta$-regularization and the \emph{Lerch's formula} \cite{iwasawa}.
Finally, by using \eqref{ratiodetpositive} and \eqref{detconfluentheun}, we get
\begin{equation}\label{detresult}
\begin{aligned}
&\mathrm{det}\mathcal{D}=\\
&\sqrt{2\,\pi}\sum_{\sigma=\pm}\frac{\Gamma\brc{-2\sigma a}\Gamma\brc{1-2\sigma a}\epsilon^{\frac{1}{2}-\mu+\sigma a}e^{-\frac{1}{2}\left(\sigma\partial_a+\partial_{a_1}+\partial_{\mu}\right)F}}{\Gamma\brc{\frac{1}{2}+\mu_1-\sigma a}\Gamma\brc{\frac{1}{2}+\mu_2-\sigma a}\Gamma\brc{\frac{1}{2}-\mu-\sigma a}}.
\end{aligned}
\end{equation}

\subsection{Confluent conformal blocks}\label{confluentblocks}

In Liouville conformal field theory, the confluent conformal blocks can be defined as collision limits of regular conformal blocks \cite{Gaiotto:2009ma, Bonelli:2011aa, Gaiotto:2012sf}. The blocks we are interested in include two primary insertions at $0$ and $1$, an irregular state of rank 1 at $\infty$, and the insertion of the degenerate field $\Phi_{2,1}$ with conformal momentum $\alpha_{2,1}=-b-\frac{1}{2b}$ at $z$. We denote these blocks as ${}_1\mathfrak{F}(z)$ following the notation of \cite{Bonelli:2022ten}.
In the semiclassical limit, in which the coupling of the theory $b\to 0$ and the momenta of the primary fields $\alpha_i\to\infty$ in such a way that their products $a_i\equiv b\,\alpha_i$ remain finite, the confluent blocks exponentiate, and the $z$-dependence becomes subleading: 
\begin{equation}\label{1F}
{}_1\mathfrak{F}(z) \sim e^{\frac{1}{b^2}F(\epsilon)+\mathcal{O}(z,b^2)}\,,
\end{equation}
where $F(\epsilon)$ is the semiclassical confluent conformal block without the degenerate insertion. The latter admits the following instanton expansion (see Appendix C.2 in \cite{Bonelli:2022ten})
\begin{equation}\label{effe}
F(\epsilon)=\frac{-1+4a^2-4a_0^2+4a_1^2}{8-32a^2}\,\epsilon+\mathcal{O}\brc{\epsilon^2}.
\end{equation}
Normalizing \eqref{1F}
with the corresponding blocks without the degenerate insertion, we obtain finite semiclassical blocks. The semiclassical limit of the BPZ equation \cite{Belavin:1984vu} 
is in this case the confluent Heun equation \eqref{confluentHeun}. As discussed in \cite{Bonelli:2022ten}, it is, therefore, possible to relate the semiclassical conformal blocks with the standard local solutions of the confluent Heun equation. We remind that the local solutions around the regular singularities are expressed in terms of the confluent Heun function
\begin{equation}
\begin{aligned}
\mathrm{HeunC}(q,\alpha,\gamma,\delta,\epsilon;z) = 1 - \frac{q}{\gamma}z + \mathcal{O}(z^2)\,,
\end{aligned}
\end{equation}
whose Taylor series expansion has a unit radius of convergence. The local solutions around the irregular singularity at infinity are, instead, defined in terms of a different function $\mathrm{HeunC}_{\infty}$, admitting an asymptotic expansion of the form
\begin{equation}
\begin{aligned}
&\mathrm{HeunC}_\infty(q,\alpha,\gamma,\delta,\epsilon;z^{-1}) \sim\\
&1+ \frac{\alpha^2 -(\gamma+\delta-1)\alpha\epsilon+(\alpha-q)\epsilon^2}{\epsilon^3} z^{-1} + \mathcal{O}(z^{-2})\,.
\end{aligned}
\end{equation}
Finally, we remind that the instanton-expansion of the composite monodromy parameter $a$ can be obtained by inverting the Matone relation \cite{Matone:1995rx}
\begin{equation}
u=\frac{1}{4}-a^2+\epsilon\,\partial_{\epsilon}F\brc{\epsilon}
\end{equation}
relating the accessory parameter of the confluent Heun equation with the semiclassical confluent conformal block.

\subsection{Additional formulas}
The coefficient $c_1(m)$ in \eqref{eqwithc1m} is given by
\begin{widetext}
\begin{equation}\label{c1m}
\begin{aligned}
c_1(m)=&\,\frac{1}{\left(4 {}_sA_{\ell m}^{(0)}-7 m^2+4 s (s+1)\right) \sqrt{4 {}_sA_{\ell m}^{(0)}-7 m^2+(2 s+1)^2}}\biggl\{{}_sA_{\ell m}^{(1)} \left[7 m^2-4 s (s+1)\right]\\
&-4 {}_sA_{\ell m}^{(0)} \left[{}_sA_{\ell m}^{(1)}-2 i \pi  M \left(7 m+2 i \sqrt{4 {}_sA_{\ell m}^{(0)}-7 m^2+(2 s+1)^2}\right) (k+i m+s)\right]\\
&-2 i \pi  M \left[[14 i m^2-8 i s (s+1)] \sqrt{4 {}_sA_{\ell m}^{(0)}-7 m^2+(2 s+1)^2}+57 m^3-4 m s (5 s+7)\right] (k+i m+s)\biggr\},
\end{aligned}
\end{equation}
\end{widetext}
where ${}_sA_{\ell m}^{(j)}$ are the coefficients of the $T_H$-expansion of ${}_sA_{\ell m}$ as in \eqref{smallTexpansion}.
In the $m=0$ case, one finds \eqref{c10}
by noticing that ${}_sA_{\ell 0}^{(0)}=\ell(\ell+1)-s(s+1)$ and ${}_sA_{\ell 0}^{(1)}=0$
because of \eqref{separationconstant}.

\subsection{Rotating BTZ case}

Let us briefly describe how the scaling in the temperature arises in the 1-loop partition function in the rotating BTZ case. The spectral problem can be written in terms of the hypergeometric equation, and the $\Gamma$-functions whose poles give the QNMs of left-moving and right-moving modes are (see for example Sec. 2.3 in \cite{Datta:2011za})
\begin{equation}\label{GammaBTZ}
\begin{aligned}
&\Gamma\left(\frac{\Delta }{2}+\frac{i l}{4 \pi  T_L}\mp \frac{ s}{2}-\frac{i \omega }{4 \pi  T_L}\right)\Gamma\left(\frac{\Delta }{2}-\frac{i l}{4 \pi  T_R}\pm\frac{ s}{2}-\frac{i \omega }{4 \pi  T_R}\right)
\end{aligned}
\end{equation}
from which
\begin{equation}
\begin{aligned}
\omega^{(L)}_n&=l-2 i \pi  T_L (\Delta +2 n\mp s),\\
\omega^{(R)}_n&=-l-2 i \pi  T_R (\Delta +2 n\pm s).
\end{aligned}
\end{equation}
The Matsubara frequencies read \cite{Castro:2017mfj}
\begin{equation}\label{MatsubaraBTZ}
\omega_k^{(M)}=l\frac{T_R-T_L}{T_L+T_R}+4 i \pi k\frac{T_L\,T_R}{T_L+T_R},
\end{equation}
where positive (negative) $k$ are associated with (anti)-QNMs.
To find the temperature scaling of the partition function as analyzed in \cite{Kapec:2024zdj}, we substitute \eqref{MatsubaraBTZ} in \eqref{GammaBTZ} and expand the right moving sector with the minus sign in front of the spin and $\Delta=s=2$ in the small $T_L$ limit (the small temperature of the Euclidean geometry corresponds to small $T_L$ and finite $T_R$):
\begin{equation}
\frac{\Delta }{2}-\frac{i l}{4 \pi  T_R}-\frac{ s}{2}-\frac{i \omega_k^{(M)} }{4 \pi  T_R}=-\frac{i l}{2 \pi  T_R}+\frac{2 \pi  k T_R+i l}{2 \pi  T_R^2}T_L+\mathcal{O}\left(T_L^2\right).
\end{equation}
In the product $\prod_{l\in\mathbb{Z}}\prod_{k\ge s}$ the modes with $l>0$ do not contribute to the temperature dependence. Therefore, by restricting to $l=0$, the selected contribution to the second $\Gamma$-function of \eqref{GammaBTZ} reduces to
\begin{equation}
\prod_{k\ge s}\Gamma\left(\frac{k\,T_L}{T_R}\right)\sim \prod_{k\ge s}\frac{T_R}{k\,T_L},
\end{equation}
which, together with the corresponding factor for anti-QNMs, produces the result $T_L^{3/2}$.

	\end{document}